\input harvmac
\def\figflag{I}


\font\blackboard=msbm10 \font\blackboards=msbm7
\font\blackboardss=msbm5
\newfam\black
\textfont\black=\blackboard
\scriptfont\black=\blackboards
\scriptscriptfont\black=\blackboardss
\def\blackb#1{{\fam\black\relax#1}}



%
\def\BC{{\blackb C}} 
 
\def\BZ{{\blackb Z}} 
\def\BP{{\blackb P}}


%
\font\mathbold=cmmib10 \font\mathbolds=cmmib7
\font\mathboldss=cmmib5
\newfam\mbold
\textfont\mbold=\mathbold
\scriptfont\mbold=\mathbolds
\scriptscriptfont\mbold=\mathboldss
\def\bi{\fam\mbold\relax}


%
\font\gothic=eufm10 \font\gothics=eufm7
\font\gothicss=eufm5
\newfam\gothi
\textfont\gothi=\gothic
\scriptfont\gothi=\gothics
\scriptscriptfont\gothi=\gothicss
\def\goth#1{{\fam\gothi\relax#1}}
%

\def\tfig#1{Fig.~\the\figno\xdef#1{Fig.~\the\figno}\global\advance\figno by1}
\def\figI{I}
%
\newdimen\tempszb \newdimen\tempszc \newdimen\tempszd \newdimen\tempsze
\ifx\figflag\figI
\input epsf
%
\def\epsfsize#1#2{\expandafter\epsfxsize{
 \tempszb=#1 \tempszd=#2 \tempsze=\epsfxsize
     \tempszc=\tempszb \divide\tempszc\tempszd
     \tempsze=\epsfysize \multiply\tempsze\tempszc
     \multiply\tempszc\tempszd \advance\tempszb-\tempszc
     \tempszc=\epsfysize
     \loop \advance\tempszb\tempszb \divide\tempszc 2
     \ifnum\tempszc>0
        \ifnum\tempszb<\tempszd\else
           \advance\tempszb-\tempszd \advance\tempsze\tempszc \fi
     \repeat
\ifnum\tempsze>\hsize\global\epsfxsize=\hsize\global\epsfysize=0pt\else\fi}}
\epsfverbosetrue
\fi
%

%
%
%
%

\def\ifigure#1#2#3#4{
\midinsert
\vbox to #4truein{\ifx\figflag\figI
\vfil\centerline{\epsfysize=#4truein\epsfbox{#3}}\fi}
\baselineskip=12pt
\narrower\narrower\noindent{\bf #1:} #2
\endinsert
}
%
%
\def\ifigures#1#2#3#4#5#6#7#8{
\midinsert
\centerline{
\hbox{\vbox{
\divide\hsize by 2
\vbox to #4truein{\ifx\figflag\figI
\vfil\centerline{\epsfysize=#4truein\epsfbox{#3}}\fi}
\baselineskip=12pt
\narrower\narrower\noindent{\bf #1:} #2
}}\qquad
\hbox{\vbox{
\divide\hsize by 2
\vbox to #8truein{\ifx\figflag\figI
\vfil\centerline{\epsfysize=#8truein\epsfbox{#7}}\fi}
\baselineskip=12pt
\narrower\narrower\noindent{\bf #5:} #6
}}}
\endinsert
}


\def\appendix#1#2{\global\meqno=1\global\subsecno=0\xdef\secsym{\hbox{#1:}}
\bigbreak\bigskip\noindent{\bf Appendix #1: #2}\message{(#1: #2)}
\writetoca{Appendix {#1:} {#2}}\par\nobreak\medskip\nobreak}

\def\ket#1{| #1 \rangle}         
%

\def\del{\partial}
\def\delb{\overline{\del}} 
\def\fourpt{\hbox{{$\rangle \kern-.25em \langle$}}} 
\def\tree{\hbox{{$\rangle \kern-.5em - \kern-.5em \langle$}}}
\def\ib{{\bar \imath}} 
\def\jb{{\bar \jmath}} %
\def\kb{{\bar k}}
\def\zb{{\bar z}}\def\wb{{\bar w}}
\def\H#1#2{{\rm H}^{#1}(#2)} 
 \def\CJ{{\cal J}}\def\CW{{\cal W}}

\def\CF{{\cal F}}

\def\ex#1{{\rm e}^{#1}}                 

\def\pd#1#2{{\partial #1\over\partial #2}}
\def\sdp{{\blackb n}}

\def\Ka{K\"ahler}
\def\cy{Calabi-Yau}
\def\LG{Landau-Ginzburg}
\def\sm{$\sigma$-model}

\def\ql{{\bi q}}
\def\qr{\overline{\bi q}}

\long\def\optional#1{}

\noblackbox

\lref\DK{J. Distler and S. Kachru, ``(0,2) Landau-Ginzburg Theory,''
{\it Nucl. Phys.} {\bf B413} (1994) 213, {\tt hep-th/9309110}.}
\lref\DKtwo{J. Distler and S. Kachru, ``Singlet Couplings and (0,2)
Models,'' {\it Nucl. Phys.} {\bf B430} (1994) 13, {\tt hep-th/9406090}.}
\lref\DKthree{J. Distler and S. Kachru, ``Quantum Symmetries and
Stringy Instantons," Phys. Lett. {\bf 336B} (1994) 368, {\tt
hep-th/9406091}.}
\lref\DKfour{J.Distler and S. Kachru, ``Duality of (0,2) String vacua",
{\tt hep-th/9501111}.}
\lref\TKawai{T. Kawai and K. Mohri, ``Geometry of (0,2) Landau-Ginzburg
Orbifolds,'' {\it Nucl. Phys.} {\bf B425} (1994) 191, {\tt hep-th/9402148}.}

\Title{\vbox{\hbox{UTTG--03--95}\hbox{\tt hep-th@xxx/9502012}}}
{\vbox{\centerline{Notes on (0,2) Superconformal Field Theories$^\star$}
}}

\centerline{Jacques Distler}\smallskip
\centerline{Theory Group}
\centerline{University of Texas at Austin}
\centerline{Austin, TX \ 78712--1081 \ USA}
\bigskip

\footnote{}{{\parindent=-5pt\par $\star$
\vtop{
\hbox{Email: {\tt distler@utpapa.ph.utexas.edu} .}
\hbox{Research supported in part by the
Robert A.~Welch Foundation, NSF Grant PHY90-09850,}
\hbox{and the A.~P.~Sloan
Foundation.}
     }     }}

\def\myab{In these lecture notes, I review the ``linear \sm" approach to (0,2)
string vacua. My aim is to provide the reader with a toolkit for studying a
very broad class of (0,2) superconformal field theories with the requisite
properties to be candidate string vacua. }\myab\ These lectures were delivered
at the 1994 Trieste Summer School.

\Date{January 1995}                 

\lref\rBott{R. Bott and L. Tu, {\it Differential Forms in Algebraic
Topology}, (Springer-Verlag, 1982).},

\lref\Sei{N. Seiberg, ``Exact Results on the Space of Vacua of
Four-Dimensional SUSY Gauge Theories,'' Rutgers preprint, {\tt
hep-th/9402044}.}
\lref\Gepner{D. Gepner, ``String Theory on Calabi-Yau Manifolds: The
Three Generation Case,'' Princeton preprint, December 1987.}
\lref\Schimmrigk{R. Schimmrigk, ``A New Construction of a
Three-Generation Calabi-Yau Manifold,'' {\it Phys. Lett.} {\bf 193B}
(1987) 175.}
\lref\Leigh{M. Dine, R.G. Leigh, and D.A. MacIntire, ``Discrete
Gauge Anomalies in String Theory,'' Santa Cruz preprint, {\tt
hep-th/9307152}. }
\lref\WitMin{E. Witten, ``On the Landau-Ginzburg Description of N=2 Minimal
Models,'' IAS preprint, {\tt hep-th/9304026}.}
\lref\Zam{A.B. Zamolodchikov, ``Irreversibility of the Flux of the
Renormalization Group in a 2d Field Theory,'' {\it JETP Lett.} {\bf 43}
(1986) 730.}

\lref\AspSmall{P. Aspinwall, B. Greene, and D. Morrison, ``Measuring
Small Distances in N=2 Sigma Models,'' IAS preprint,
{\tt hep-th/9311042}.}
\lref\MonomialDivisor{P. Aspinwall, B. Greene and D. Morrison, ``The Monomial
Divisor Mirror Map,"  {\tt alg-geom/9309007.} }
\lref\MorrisonPlesser{D. Morrison and R. Plesser, ``Summing the Instantons:
Quantum Cohomology and Mirror Symmetry in Toric Varieties," DUKE-TH-94-78,
{\tt hep-th/9412236}.}

\lref\SpecialGeo{B. de Wit, P. Lauwers and A. van Pr\oe yen, Nucl. Phys. {\bf
B255} (1985) 569.}
\lref\ranspec{S. Cecotti, S. Ferrara and L. Girardello, Int. J. Mod. Phys.
{\bf A4} (1989) 2475\semi
L. Castellani, R. D'Auria and S. Ferrara, Class. Quantum Grav. {\bf 7} (1990)
1767.}
\lref\CandMod{P. Candelas and X. De la Ossa, ``Moduli
Space of Calabi-Yau Manifolds,'' {\it Nucl. Phys.} {\bf B355} (1991)
455.}
\lref\OldEd{E. Witten, ``Symmetry Breaking Patterns in Superstring
Models,'' {\it Nucl. Phys.} {\bf B258} (1985) 75.}
\lref\DKL{L. Dixon, V. Kaplunovsky and J. Louis, Nucl. Phys. {\bf B329}
(1990) 27.}
 \lref\StromSpec{A. Strominger, ``Special Geometry,''
 {\it Comm. Math. Phys.} {\bf 133}
(1990) 163.}
\lref\PerStrom{V. Periwal and A. Strominger, Phys. Lett. {\bf B335} (1990)
261.}
\lref\TopAntiTop{S. Cecotti and C. Vafa, Nucl. Phys. {\bf B367} (1991) 359.}
\lref\AandB{E. Witten, in ``Proceedings of the Conference on Mirror Symmetry",
MSRI (1991).}
\lref\LVW{W. Lerche, C. Vafa and N. Warner, Nucl. Phys. {\bf B324} (1989)
427.} 
\lref\twisted{E. Witten. Comm. Math. Phys. {\bf 118} (1988) 411\semi
E. Witten, Nucl. Phys. {\bf B340} (1990) 281\semi
T. Eguchi and S.-K. Yang , Mod. Phys. Lett. {\bf A5} (1990) 1693.}
 \lref\GVW{B. Greene, C. Vafa and N. Warner,
``Calabi-Yau Manifolds and Renormalization Group Flows,''
 {\it Nucl. Phys.} {\bf B324} (1989)
371.}
\lref\Grisaru{M. Grisaru, A. Van de Ven and D. Zanon, Phys. Lett. {\bf 173B}
(1986) 423.}
\lref\SeiNat{N. Seiberg, ``Naturalness Versus Supersymmetric
Non-renormalization Theorems,'' {\it Phys. Lett.} {\bf B318} (1993)
469, {\tt hep-ph/9309335}.}

\lref\DSWW {M. Dine, N. Seiberg, X.G. Wen and E. Witten,
``Non-Perturbative Effects on the String World Sheet I,'' {\it Nucl.
Phys.}~{\bf B278} (1986) 769, ``Non-Perturbative Effects on the String
World Sheet II,'' {\it Nucl. Phys.}~{\bf B289} (1987) 319. }
\lref\DixonRev{L. Dixon, in ``Proceedings of the 1987 ICTP Summer Workshop in
High Energy Physics and Cosmology", ed G. Furlan, {\it et. al.}}
\lref\Exact{J. Distler and B. Greene,
``Some Exact Results on the Superpotential from Calabi-Yau
Compactifications,''
{\it Nucl. Phys.}
 {\bf B309} (1988) 295.}
\lref\twozero{J. Distler and B. Greene,
``Aspects of (2,0) String Compactifications,''
{\it Nucl. Phys.} {\bf B304} (1988) 1.}
\lref\AspMor{P. Aspinwall and D. Morrison, ``Topological Field Theory and
Rational Curves," Comm. Math. Phys. {\bf 151} (1993) 245.}
\lref\CandMir{P. Candelas, X. de la Ossa, P. Green and L. Parkes,
``A Pair of Calabi-Yau Manifolds as an Exactly Soluble Superconformal
Theory,''
{\it Nucl.
Phys.} {\bf B359} (1991) 21.}
\lref\Kutconf{D. Kutasov, ``Geometry on the space of conformal field
theories and contact terms,'' Phys. Lett. {\bf B220} (1989) 153.}%
\lref\GrSei{M. Green and N. Seiberg, ``Contact interactions in
superstring theory," Nucl. Phys. {\bf B299} (1988) 559.}%
\lref\WilZee{F. Wilczek and A. Zee, Phys. Rev. Lett. {\bf 52} (1984) 2111.}
\lref\Banks{T. Banks, L. Dixon,
D.Friedan and E. Martinec, ``Phenomenology and Conformal Field Theory or
Can String Theory Predict the Weak Mixing Angle?,'' {\it Nucl. Phys.}
{\bf B299} (1988) 613.}
\lref\Sen{A. Sen, Nucl. Phys. {\bf B278} (1986) 289.}
\lref\Hodge{P. Griffiths, Periods of Integrals on Algebraic manifolds I,II,
Am. J. Math. {\bf 90} (1970) 568,805\semi
R. Bryant and P. Griffiths, in ``Progress in Mathematics {\bf 36}"
(Birkh\"auser, 1983) 77.}
\lref\Odake{S. Odake, Mod. Phys. Lett. {\bf A4} (1989) 557\semi
S. Odake, Int. Jour. Mod. Phys. {\bf A5} (1990) 897\semi
T. Eguchi, H. Ooguri, A. Taormina and S-K. Yang, Nucl. Phys. {\bf B315}
(1989) 193.}
\lref\GSW{M. Green, J. Schwarz and E. Witten, ``Superstring theory, vol. II"
(Cambridge University Press, 1987).}
\lref\CHSW{P. Candelas, G. Horowitz, A. Strominger and E. Witten,
``Vacuum Configurations for Superstrings,'' {\it Nucl. Phys.} {\bf
B258} (1985) 46.}
\lref\NR{L. Alvarez-Gaum\'e, S. Coleman and P. Ginsparg, Comm. Math. Phys.
{\bf 103} (1986) 423.}
\lref\manin{D. Leites, ``Introduction to the theory of supermanifolds", Russ.
Math. Surveys {\bf 35} (1980) 3\semi
Yu. Manin, ``Gauge field theory and complex geometry", (Springer, 1988).}
\lref\Reviews{
J. Schwarz, ``Superconformal symmetry in string theory", lectures
at the 1988 Banff Summer Institute on Particles and Fields (1988)\semi
D. Gepner, ``Lectures on N=2 string theory",
lectures at the 1989 Trieste Spring School (1989)\semi
N. Warner, ``Lectures on N=2 superconformal theories and singularity theory",
lectures at the 1989 Trieste Spring School (1989)\semi
B. Greene, ``Lectures on string theory in four dimensions",
lectures at the 1990 Trieste Spring School (1990)\semi
S. Yau (editor), ``Essays in Mirror Manifolds",  Proceedings of the Conference
on Mirror Symmetry, MSRI (International Press, 1992).
}
\lref\Trieste{J. Distler, ``Notes on N=2 $\sigma$-models," lectures at the
1992 Trieste Spring School (1992), {\tt hep-th/9212062.}}
\lref\Seiberg{N. Seiberg, Nucl. Phys. {\bf B303} (1988) 286.}
\lref\Dubrovin{B. Dubrovin, ``Geometry and integrability of
topological--antitopological fusion", INFN preprint, INFN-8-92-DSF (1992).}
\lref\GPM{B. Greene, D. Morrison, and R. Plesser, in preparation.}
\lref\GPmirror{B. Greene and R. Plesser, Nucl. Phys. {\bf B338} (1990) 15.}
\lref\phases{E. Witten, ``Phases of N=2 Theories in Two Dimensions,"
{\it Nucl. Phys.} {\bf B403} (1993) 159, {\tt hep-th/9301042}.}
\lref\Vafa{C. Vafa, ``String Vacua and Orbifoldized LG models,''
{\it Mod. Phys. Lett.} {\bf A4} (1989) 1169.}
\lref\Ken{K. Intriligator and C. Vafa, ``Landau-Ginzburg Orbifolds,''
{\it Nucl. Phys.} {\bf B339} (1990) 95.}
\lref\Us{S. Kachru and E. Witten, ``Computing The Complete Massless
Spectrum Of A Landau-Ginzburg Orbifold,''
{\it Nucl. Phys.} {\bf B407} (1993) 637, {\tt hep-th/9307038}.}
\lref\WitMin{E. Witten, ``On the Landau-Ginzburg Description of N=2
Minimal Models,'' IAS preprint
IASSNS-HEP-93/10, {\tt hep-th/9304026}.}
\lref\Fre{P. Fr\'e, F. Gliozzi, M. Monteiro, and A. Piras, ``A
Moduli-Dependent Lagrangian For (2,2) Theories On Calabi-Yau n-Folds,''
{\it Class. Quant. Grav.} {\bf 8} (1991) 1455; P. Fr\'e, L. Girardello,
A. Lerda, and P. Soriani, ``Topological First-Order Systems With
Landau-Ginzburg Interactions,'' {\it Nucl. Phys.} {\bf B387} (1992)
333, {\tt hep-th/9204041}.}
\lref\Greene{B.R. Greene, ``Superconformal Compactifications in Weighted
Projective Space,'' {\it Comm. Math. Phys.} {\bf 130} (1990) 335.}
\lref\fendley{P. Fendley and K. Intriligator, ``Central Charges
Without Finite Size Effects,'' Rutgers preprint RU-93-26, {\tt
hep-th/9307101}.}
\lref\OldWit{E. Witten, ``New Issues in Manifolds of SU(3) Holonomy,''
 {\it Nucl. Phys.} {\bf B268} (1986) 79.}
\lref\Pasquinu{S. Cecotti, L. Girardello, and A. Pasquinucci,
``Non-perturbative Aspects and Exact Results for the N=2 Landau-Ginzburg
Models,'' {\it Nucl. Phys.}~{\bf B338} (1989) 701, ``Singularity
Theory and N=2 Supersymmetry,'' {\it Int. J. Mod. Phys.} {\bf A6} (1991)
2427.}
\lref\KT{A. Klemm and S. Theisen, ``Mirror Maps and Instanton Sums for
Complete Intersections in Weighted Projective Space,'' Preprint LMU-TPW
93-08, {\tt hep-th/9304034}. }
\lref\VafaQ{C. Vafa, ``Quantum Symmetries of String Vacua,''{\it Mod. Phys.
Lett.} {\bf A4} (1989) 1615. }
\lref\HWPmirrors{D. Morrison, ``Picard-Fuchs Equations and Mirror Maps for
Hypersurfaces," In {\it Essays on Mirror Manifolds}, ed. S.--T. Yau,
(Int. Press Co., 1992) {\tt alg-geom/9202026}\semi
A. Font, ``Periods and Duality Symmetries in Calabi-Yau
Compactifications,'' {\it Nucl. Phys.} {\bf B391} (1993) 358, {\tt
hep-th/9203084}\semi
A. Klemm and S. Theisen, ``Considerations of One Modulus Calabi-Yau
Compactification: Picard-Fuchs Equation, K\"ahler Potentials and Mirror
Maps,"
{\it Nucl. Phys.} {\bf B389} (1993) 153, {\tt hep-th/9205041}.}
\lref\flops{P. Aspinwall, B. Greene and D. Morrison, ``Multiple Mirror
Manifolds and Topology Change in String Theory," {\it Phys. Lett.} {\bf 303B}
(1993) 249, {\tt hep-th/9301043}.}
\lref\flopsII{P. Aspinwall, B. Greene and D. Morrison, ``Calabi-Yau Moduli
Space, Mirror Manifolds and Spacetime Topology Change in String Theory,"
IAS and Cornell preprints IASSNS-HEP-93/38, CLNS-93/1236, to appear.}
\lref\cvetic{M. Cvetic, ``Exact Construction of (0,2) Calabi-Yau
Manifolds,'' {\it Phys. Rev. Lett.} {\bf 59} (1987) 2829.}
\lref\Miron{J. Distler, B. Greene, K. Kirklin and P. Miron, ``Calculating
Endomorphism Valued Cohomology: singlet spectrum in superstring models,"
{\it Comm. Math. Phys.} {\bf 122} (1989) 117.}
\lref\miracles{M. Dine and N. Seiberg, ``Are (0,2) Models String Miracles?,"
{\it Nucl. Phys.} {\bf B306} (1988) 137.}
\lref\GrPl{B.R. Greene and M.R. Plesser, ``Mirror Manifolds: A Brief
Review and Progress Report,'' {\tt hep-th/9110014}.}
\lref\masses{P. Candelas, X. De la Ossa and collaborators, to appear.}
\lref\eva{E. Silverstein and E. Witten, ``Global U(1) R-Symmetry and Conformal
Invariance of (0,2) Models," IASSNS-94/4,PUPT-1453, {\tt hep-th/9403054}.}
\lref\WittenElGen{E. Witten, ``Elliptic Genera and Quantum Field
Theory,'' {\it Comm. Math. Phys.} {\bf 109} (1987) 525\semi
``The Index of the Dirac Operator in Loop Space,'' in {\it Elliptic
Curves and Modular Forms in Algebraic Topology}, P.S. Landweber ed.,
Lecture Notes in Mathematics 1326 (Springer-Verlag, 1988).}

\lref\Mohri{T. Kawai and K. Mohri, ``Geometry of (0,2) Landau-Ginzburg
Orbifolds,'' KEK preprint, {\tt hep-th/9402148}.}
\lref\Nemeschansky{D. Nemeschansky and N. Warner, ``Refining the
Elliptic Genus,'' USC preprint, {\tt hep-th/9403047}.}
\lref\Morse{C. Vafa, ``c-theorem and the Topology of 2d QFTs,'' {\it Phys.
Lett.}
{\bf 212B} (1988) 28 \semi
S. Das, G. Mandal, and S. Wadia, ``Stochastic Differential Equations on
Two-Dimensional Theory Space and Morse Theory,'' {\it Mod. Phys. Lett.}
{\bf A4} (1989) 745.}
\lref\BottHomotopy{R. Bott, `` Nondegenerate Critical Manifolds,"
{\it Ann. Math.} {\bf 60} (1954) 248\semi
R. Bott, ``The Stable Homotopy of the Classical Groups,"
{\it Ann. Math.} {\bf 70} (1959) 313.}
\lref\rgrassmannian{E. Witten, ``The Verlinde Algebra and the Cohomology of the
Grassmannian," IASSNSS-HEP-93-41, {\tt hep-th/9312104}.}
\lref\KazamaSuzuki{Y. Kazama and H. Suzuki, ``New N=2 Superconformal Field
Theories and Superstring Compactification," Nucl. Phys. {\bf B321} (1989) 232.
}

\lref\ShifVain{M. Shifman and A. Vainshtein, Nucl. Phys. {\bf B277} (1986)
456\semi Nucl. Phys. {\bf B359} (1991) 571.}
\lref\SeibWit{N. Seiberg and E. Witten,``Monopole
Condensation, And Confinement in $N=2$ Supersymmetric Yang-Mills Theory,"
 Nucl. Phys. {\it B426} (1994) 19, {\tt hep-th/9407087}.}
\lref\BanksDine{T. Banks and M. Dine, ``Coping with strongly coupled
string theory," Phys. Rev. {\bf D50} (1994) 7454, {\tt hep-th/9406132.}}
\lref\DineHolo{M. Dine and Y. Shirman, ``Some Explorations in Holomorphy,"
Phys. Rev. {\bf D50} (1994) 5389, {\tt hep-th/9405155}.}

\lref\levelmatching{C. Vafa, ``Modular Invariance and Discrete Torsion on
Orbifolds,'' {\it Nucl. Phys.} {\bf B273} (1986) 592.}

\centerline{\bf Notes on (0,2) Superconformal Field Theories$^\star$}
\smallskip
{\baselineskip=14pt
\centerline{Jacques Distler}\smallskip
\centerline{Theory Group}
\centerline{University of Texas at Austin}
\centerline{Austin, TX \ 78712--1081 \ USA}
\bigskip\bigskip

\footnote{}{{\parindent=-5pt\par $\star$
\vtop{
\hbox{Email: {\tt distler@utpapa.ph.utexas.edu} .}
\hbox{Research supported in part by the
Robert A.~Welch Foundation, NSF Grant}
\hbox{PHY90-09850, and the A.~P.~Sloan Foundation.}
}     }}
\parindent=20pt

\narrower\narrower\noindent
\myab

}

\newsec{Invitation au voyage}

The case for considering string compactifications with spacetime
supersymmetry is often made in four dimensional terms, say, that this
provides a solution to the hierarchy problem. But the real justification is
that these are the {\it only} known perturbative solutions to the fermionic
string with four flat noncompact directions. So if perturbative
string theory can tell us anything, these are the starting points about which
to do our perturbation theory. Now, it is clear that not all of the physics
is captured by perturbation theory. However, {\it some} of it is (and we
even have reason \BanksDine\ to believe that many of the crucial features are)
captured by perturbation theory. Besides, you have to walk before you can run
$\dots$.

Spacetime supersymmetry is equivalent to (0,2)
superconformal symmetry on the worldsheet
with  an integrality condition on the
right-moving
$U(1)$ charges (see \refs{\DixonRev,\Banks,\Sen}). So if we want to
study perturbative solutions to string theory, we should be studying (0,2)
SCFTs. Surprisingly little is known about the subject, given its evident
importance. Almost all of the results we have to date are for the very
special case of left-right symmetric $(2,2)$ superconformal field
theories, or very simple orbifolds thereof. This is clearly a very special case
of a (0,2) SCFT, where we simply ignore the left-moving superconformal
symmetry. For a long time, more general (0,2) SCFTs were considered simply too
hard to study. I hope in these lectures to convince you that they are not too
hard, and that, because there are so many open questions, this is fertile area
to work on.

Mainly in these lectures, I will concentrate on giving you a ``toolkit" for
building and studying  (0,2) models of the requisite sort. More details can
be found in the references, as can some results that one can learn both about
the general features of (0,2) theories and the specifics of particular models.
Much of the work that I describe in these notes was done jointly with
S.~Kachru.

\newsec{Generalities}
As I said, $N=1$ Spacetime supersymmetry is  equivalent
\refs{\DixonRev,\Banks,\Sen}  to (0,2) superconformal symmetry
on the worldsheet, provided a certain integrality condition on the $U(1)$
charges holds. The origin of this condition is that it is required so that we
may define a chiral GSO projection. You might think, therefore,
 that what we
will look for is a theory whose symmetry algebra is
$(vir)_L\times(N=2\ svir)_R$.
In fact, we will require a somewhat larger symmetry. Namely, we will require
as well that there exist a left-moving $U(1)$ current algebra of level $r$,
that
is,
$$J(z)J(w)= {r\over (z-w)^2}$$
and the symmetry algebra will be
$(\widehat{U(1)}\sdp vir)_L\times(N=2\ svir)_R$, with $(c,\bar c)=(6+r,9)$.

To turn this into a string theory, we add four free bosons $X^\mu$, and their
(0,1) superpartners, four free Majorana-Weyl fermions $\psi^\mu$. We also add
$\lambda^I,\ I=1,\dots 16-2r$, free left-moving  Majorana-Weyl fermions which
yield a linearly-realized $SO(16-2r)$ subgroup of the spacetime gauge group,
and we add a left-moving $E_8$ current algebra.

The left-moving $\widehat{U(1)}$ current algebra plays a dual role in the
theory. First, it provides another linearly-realized piece of the spacetime
gauge group (which, at this stage, appears to be $SO(16-2r)\times U(1)\times
E_8$ ). Second, it provides a candidate for a chiral GSO projection for the
left-movers:
\eqn\gsodef{g=\ex{-i\pi J_0}(-1)^{F_{\lambda^I}}}
where $F_{\lambda^I}$ is the fermion number for the left-moving free fermions.

In keeping with its role in forming the GSO projection, the left-moving
$\widehat{U(1)}$ also provides the left-moving spectral flow generator
(ground state of the left-moving Ramond sector\foot{In the
Ramond sector, $J_0$ is replaced by $J_0 +r/2$ in the formula for the GSO
projection.}) which
promotes the spacetime gauge group to $E_6$, $SO(10)$, or $SU(5)$, for
$r=3,4,5$.

We are all familiar \GSW\
with how the representations of $SO(10)\times U(1)$ assemble
themselves into representations of $E_6$ in (2,2) compactifications ($r=3$).
The situation for $r=4,5$ may be more unfamiliar, so I have summarized it in
the following tables.
\def\tablerule{\omit&\multispan{6}{\tabskip=0pt\hrulefill}&\cr}
\def\tablepad{\omit&height3pt&&&&&&&\cr}
$$\vbox{\offinterlineskip\tabskip=0pt\halign{
\strut$#$\quad&\vrule#&\quad\hfil $#$ \hfil\quad &\vrule #&\quad \hfil $#$
\hfil \quad&\vrule #& \quad $#$ \hfil\ &\vrule#&\quad $#$\cr
&\omit&\hbox{Rep.~of $SO(10)$}&\omit&\hbox{Rep.~of $SO(8)\times
U(1)$}&\omit&\hbox{Cohomology Group}&\omit&\cr
\tablerule\tablepad
&&45&&8^{s'}_{-2}\oplus(28_0\oplus1_0)\oplus 8^{s'}_2&&\H{*}{M,\CO}&&\cr
\tablepad\tablerule\tablepad
 r=4&&16&&8^{s}_{-1}\oplus8^{v}_1&&\H{*}{M,V}&&\cr
\tablepad\tablerule\tablepad
&&10&&1_{-2}\oplus8^{s'}_{0}\oplus1_2&&\H{*}{M,\bigwedge^2V}&&\cr
\tablepad\tablerule\tablepad
&&1&&1_0&&\H{*}{M,End\ V}&&\cr \tablepad\tablerule
\noalign{\bigskip}
&\omit&\hbox{Rep.~of $SU(5)$}&\omit&\hbox{Rep.~of $SO(6)\times
U(1)$}&\omit&\hbox{Cohomology Group}&\omit&\cr
\tablerule\tablepad
&&24&&\bar 4_{-5/2}\oplus(15_0\oplus1_0)\oplus 4_{5/2}&&\H{*}{M,\CO}&&\cr
\tablepad\tablerule\tablepad
 r=5&&10&&4_{-3/2}\oplus6_1&&\H{*}{M,V}&&\cr
\tablepad\tablerule\tablepad
&&\bar 5&&\bar 4_{-1/2}\oplus 1_2&&\H{*}{M,\bigwedge^2V}&&\cr
\tablepad\tablerule\tablepad
&&1&&1_0&&\H{*}{M,End\ V}&&\cr \tablepad\tablerule
\noalign{\bigskip}
\noalign{\narrower\noindent{\bf Table 1:} Representations of the
linearly realized part of the gauge group and how they assemble themselves.}
 }}$$

Note that the representations which appear alternate between spinor and tensor
representations of $SO(16-2r)$, and the $U(1)$ charge jumps by $r/2$ with
each application of the spectral flow.
One realization of this general setup is (2,2) superconformal field theory.
In this case, $r=3$, since the left-moving $N=2$ superconformal algebra with
$c=3r$ contains
a $\widehat{U(1)}$ subalgebra at level $r$. Clearly, though, this is a very
special case. Phenomenologically, it may also be a relatively unattractive one,
as $r=4,5$ seems to lead to more attractive phenomenology.

\newsec{Nonlinear $\sigma$ Models}

The (2,2) nonlinear sigma model can be written
\eqn\nlsm{\eqalign{
S={i\over 2\pi}\int &\half g_{i\jb}(\del X^i\delb X^\jb+\del X^\jb\delb X^i)
-\half b_{i\jb}(\del X^i\delb X^\jb-\del X^\jb\delb X^i)\cr
&+i(\psi_\ib D\psi^\ib +\lambda_i \bar D \lambda^i)
+ R^k{}_l{}^\ib{}_\jb(X)\lambda_k\lambda^l\psi_\ib\psi^\jb\cr
}}
where $X:\Sigma\to M$ is the \sm\ map from the worldsheet into a \Ka\ manifold
$M$, which for reasons that will become clear shortly, we will assume has
vanishing first Chern class, $c_1(T)=0$. The left- and right-moving fermions
couple to the appropriate pullback connections,
$D\psi^\ib=\del\psi^\ib+\del X^\jb\Gamma^\ib_{\jb\kb}(X)\psi^\kb$, {\it etc.}

The (0,2) generalization of this is to replace the action for the left-moving
fermions by
\eqn\nlsmzt{S={i\over 2\pi}\int \dots +i(\dots+\lambda_a\bar D\lambda^a)
+ F^a{}_b{}^\ib{}_\jb(X)\lambda_a\lambda^b\psi_\ib\psi^\jb}
where now the $\lambda^a$ transform as sections of a holomorphic vector bundle
$V\to M$ with
\eqn\eCconds{c_1(V)=0,\qquad c_2(V)=c_2(T)}
The data specifying the \sm\ now is: the \Ka\ metric, $g_{i\jb}(X)$, a closed
2-form $b_{i\jb}(X)$, and the holomorphic connection on $V$, $A^a{}_{bi}(X)$,
whose curvature is $F^a{}_{bi\jb}(X)$.

For string theory, of course, we are interested in a {\it
conformally invariant} \sm. Requiring conformal invariance imposes some
conditions on the above data. For instance, demanding that the 1-loop
$\beta$-function of \nlsmzt\ vanish requires that $g_{i\jb}$ be Ricci-flat.
However, these conditions are corrected at higher orders in \sm\ perturbation
theory, and we don't have the slightest idea what the ``all orders" equation
necessary for conformal invariance is.

In the face of this obstacle, there are two attitudes one can adopt. The
first is to imagine that we can construct the exact conformally invariant
theory order by order in perturbation theory, starting with a solution to the
1-loop
$\beta$-function equations. If the \sm\ is weakly coupled, we might expect
that the exact conformally invariant theory is ``close" to its 1-loop
approximation.

A more fruitful point of view is to accept that the \sm\ \nlsmzt\ (or, at
least any \sm\ we can actually write down) is not conformally-invariant.
However, it flows under the Renormalization Group to an infrared fixed
point theory which is the desired conformally invariant theory.

This second point of view is very useful. It suggest several helpful ways of
looking at the \sm. First, the RG flow is dissipative. The data
$g_{i\jb},b_{i\jb},A^a{}_{bi}$ represent an infinite number of coupling
constants in the two dimensional quantum field theory. All but a finite
number of these are marginally irrelevant and flow to zero in the infrared.
Thus the fixed-point theory is characterized by a finite number of parameters
which are RG-invariant.

So what are the RG-invariant parameters characterizing \nlsmzt?
They are
\item{a)}the complex structure of $M$
\item{b)}the holomorphic structure of the vector bundle $V$
\item{c)}the cohomology class of the complex \Ka\ form $\CJ=B+iJ$, where
$$J=i\ g_{i\jb}dX^i\wedge dX^\jb,\qquad B=\ b_{i\jb}dX^i\wedge dX^\jb$$

The first two are automatic in this formalism. They are assured by the
existence of a chiral $U(1)_L\times U(1)_R$ symmetry under which
\eqn\echiralsym{\eqalign{
\psi^\ib&\to\ex{i\theta_R}\psi^\ib,\qquad
\lambda^a\to\ex{-i\theta_L}\lambda^a\cr
\psi_\ib&\to\ex{-i\theta_R}\psi_\ib,\qquad
\lambda_a\to\ex{i\theta_L}\lambda_a\cr
 }}
Note that the conditions $c_1(T)=c_1(V)=0$ are precisely what is needed
to ensure that this chiral $U(1)_L\times U(1)_R$ is nonanomalous. In the
conformal limit, the corresponding conserved $U(1)$ currents become the
generators of the left-moving $\widehat{U(1)}$ current algebra and the
right-moving $\widehat{U(1)}_R$ current algebra in the N=2 superconformal
algebra.

The RG-invariance of the cohomology class of $\CJ$ is, by contrast, highly
nontrivial. It was proven to all orders in perturbation theory in \NR, where
it was shown that all perturbative corrections to $J$ are exact two-forms.
Beyond perturbation theory, one needs to worry about \sm\ instantons,
topologically nontrivial maps from the worldsheet into $M$. Naively,
corrections to $g_{i\jb}$ are instanton-antiinstanton effects, and so rather
hard to see. There are rather indirect arguments \twozero\ which one might use
to try to show that the cohomology class of $\CJ$ is unrenormalized, even when
\sm\ instantons are taken into account. But the necessary conditions are very
hard to verify, and for a long time this pretty much stymied any progress on
(0,2) \sm s.

Since nonlinear \sm s are so hard, we can invoke another great principle of
the renormalization group, namely {\it universality}. There are many QFTs
which renormalize to the same IR fixed point. If nonlinear \sm s are too
hard, we should look for another, simpler family of QFTs which happen to be in
the same universality class. This motivates us to look at {\it linear} \sm s
\phases.

\newsec{Linear \sm s}

Since we will be interested in (0,2) linear \sm s, we should first discuss
(0,2) superfields. Our (0,2) superspace has coordinates
$(z,\zb,\theta^+,\theta^-)$. The spinor derivatives are
$$\bar D_\pm=\pd{}{\theta^\pm}+\theta^\mp\partial_\zb$$
Chiral (scalar) superfields $\Phi$ satisfy
$$\bar D_+\Phi=0$$
In components,
$$\Phi=\phi+\theta^-\psi+\theta^-\theta^+\partial_\zb\phi$$
A (chiral) fermi superfield $\Lambda$ also satisfies the chiral constraint
$\bar D_+\Lambda=0$, but its lowest component is a left-handed fermion
$\lambda$, and its upper component is an auxiliary field $l$:
$$\Lambda=\lambda+\theta^-l+\theta^-\theta^+\partial_\zb \lambda$$

The (0,2) gauge multiplet actually consists of a pair of (0,2) superfields
$\CA,V$, where $V$ is a superfield whose Minkowski continuation is a {\it
real} superfield, and $\CA$ is one whose Minkowski continuation is pure {\it
imaginary}. The lowest component of $V$ is a real scalar, and the lowest
component of $\CA$ is the left-moving component of the gauge field, $a$
(which we take to be anti-Hermitian).

Super-gauge transformations act on $\CA,V$ as
$$V\to V-i(\chi-\bar\chi),\qquad\CA\to\CA-i(\chi+\bar\chi)$$
where $\chi$ is a chiral scalar superfield, $\bar D_+\chi=\bar D_-\bar\chi=0$.

In Wess-Zumino gauge, the nonzero components of the gauge multiplet are
$$\eqalign{V&=\theta^-\theta^+\bar a\cr
\CA&=a+\theta^+\alpha-\theta^-\bar \alpha+\half\theta^-\theta^+ D\cr
}$$
(The residual gauge symmetry in WZ gauge is
$\chi=\rho+\theta^-\theta^+\partial_\zb\rho$, with $\rho$ real.) $a,\bar a$
are the left- and right-moving components of the gauge field,
$\alpha,\bar\alpha$ are the left-moving gauginos, and $D$ is a (real)
auxiliary field.

Under a super gauge transformation,
$$\Phi\to \ex{2iQ\chi}\Phi,\qquad\bar\Phi\to\ex{-2iQ\bar \chi}\bar\Phi$$
where $Q$ is the charge of $\Phi$, and similarly for $\Lambda$.
Let
$$\tilde\Phi=\ex{QV}\Phi,\qquad\tilde{\bar\Phi}=\ex{QV}\bar\Phi$$
and similarly for $\Lambda$.

A gauge invariant kinetic term for $\Phi$ is
\eqn\eSphi{\eqalign{S_\Phi&=\int
d^2zd^2\theta\ (\partial_z-Q\CA)\tilde{\bar\Phi}\tilde\Phi-
\tilde{\bar\Phi}(\partial_z+Q\CA)\tilde\Phi\cr
&=\int d^2z\ (\partial_z-Qa)\bar\phi(\partial_\zb+Q\bar a)\phi+
(\partial_\zb-Q\bar a)\bar\phi(\partial_z+Qa)\phi\cr
&\qquad
+2\bar\psi(\partial_z+Qa)\psi
+Q(\bar\alpha\bar\psi\phi-\alpha\psi\bar\phi)- Q\bar\phi\phi D\cr }}
An invariant kinetic term for $\Lambda$ is
\eqn\eSlambda{\eqalign{
S_\lambda&=\int d^2zd^2\theta\ \tilde{\bar\Lambda}\tilde\Lambda\cr
&=\int d^2z\ 2\bar\lambda(\partial_\zb+Q\bar a)\lambda-\bar l l\cr}}

Define the spinor covariant derivatives
$$\bar\CD_\pm=\pm\ex{\pm V}\bar D_\pm\ex{\mp V},\qquad \CD=\partial_z+\CA
$$
and the corresponding gauge field strengths
$$\eqalign{
\CF&=2[\CD,\bar\CD_+]
=-\alpha+\theta^-(D+f)-\theta^-\theta^+\partial_\zb\alpha\cr
\bar\CF&=2[\CD,\bar\CD_-]
=-\bar\alpha+\theta^+(D-f)+\theta^-\theta^+\partial_\zb\bar\alpha\cr
}$$
where $f=2(\partial_z \bar a -\partial_\zb a)$\ \foot{The normalization is such
that $d^2z={i\over 2}dz\wedge d\zb$, so ${1\over 2\pi}\int d^2z\
f=n_{inst}\in\BZ$.}. Note that
$\CF$ is chiral:
$\bar D_+\CF=0$. The kinetic term for the gauge fields is
\eqn\eSgauge{\eqalign{S_{gauge}&=-{1\over2 e^2}\int
d^2zd^2\theta\CF\bar\CF\cr
&={1\over2 e^2}\int d^2 z( f^2-
D^2+2\alpha\partial_\zb\bar\alpha)\cr}}
and the Fayet-Iliopoulos D-term is
\eqn\eSD{\eqalign{
S_D&=-{it\over 2}\int d^2z d\theta^-\CF+{i\bar t\over 2}\int
d^2z d\theta^+\bar \CF\cr
&=r\int d^2z D-{i\theta\over 2\pi}\int d^2z f\cr}}
where $t={\theta\over 2\pi}+ir$.

Finally, a (0,2) superpotential is
\eqn\eSW{\eqalign{
S_\CW&=\int d^2d\theta^-\ \goth{m}\Lambda F(\Phi)+\int
d^2d\theta^+\ \bar \goth{m}\overline{\Lambda}\ \overline{F(\Phi)}\cr
&=\int d^2z\ \goth{m}(lF(\phi)-\lambda\pd{F}{\phi}\psi)+{\rm h.c.}\cr }}
where $F$ is a homogeneous polynomial of the appropriate degree such that
\eSW\ is
gauge invariant, and $\goth{m}$ is a coupling constant with dimensions of mass.
It is commonplace to set $\goth{m}=1$,  which simplifies the notation, but it
is useful to remember that it is there.

This is all that is required in order to discuss (0,2)
supersymmetric linear
\sm s. But if we wish to discuss (2,2)  supersymmetric  theories (and certain
(0,2) generalizations), we actually need to enlarge the gauge multiplet.
we introduce a complex fermionic superfield $\Sigma$ and its conjugate
$\bar\Sigma$. N.B.~ these do {\it not} obey a chiral constraint!
Correspondingly, we introduce a new {\it gauge} symmetry, under which
\eqn\enewgauge{
\matrix{\Sigma\to\Sigma+i\Omega&\bar\Sigma\to\bar\Sigma-i\bar\Omega\cr
\Lambda\to\Lambda
+2iQ\Omega\Phi&\bar\Lambda\to\bar\Lambda-2iQ\bar\Omega\bar\Phi\cr}}
with $\Omega$ a chiral fermionic superfield, and all other fields being
invariant.

$\Sigma$ has four independent components, but the $\Omega$ gauge symmetry
allows us to gauge two of them away. We'll call the ones that remain
$\sigma,\beta$, and note that they appear in the gauge invariant quantity
$$\bar
D_+\Sigma=\half(\sigma+\theta^-\beta+\theta^-\theta^+\partial_\zb\sigma)$$
Unfortunately, the action \eSlambda\ is not invariant under the $\Omega$ gauge
symmetry. To correct this, we add
\eqna\eSsigma
$$\eqalignno{S_\Sigma&=\int d^2zd^2\theta\
Q^2\tilde{\bar\Phi}\tilde\Phi\bar\Sigma\Sigma
-Q(\tilde{\bar\Lambda}
\tilde\Phi\Sigma-\tilde{\bar\Phi}\tilde\Lambda\bar\Sigma)&\eSsigma
a\cr
&+{1\over 2e^2}(-\bar D_-\bar\Sigma\partial_z\bar D_+\Sigma+\partial_z\bar
D_-\bar\Sigma\bar D_+\Sigma)&\eSsigma b\cr}$$
$\eSsigma{a}$ makes \eSlambda\ invariant, while $\eSsigma{b}$ gives $\Sigma$
a kinetic term. In ``Wess-Zumino" gauge,
\eqn\eSsigmacomp{\eqalign{
S_\Sigma&=\int d^2z\
Q^2|\phi|^2|\sigma|^2+Q(\bar\lambda\psi\sigma-\lambda\bar\psi\bar\sigma)
-Q(\beta\bar\lambda\phi-\bar\beta\lambda\bar\phi)\cr
&\qquad +{1\over
2e^2}(\partial_z\bar\sigma\partial_\zb\sigma+
\partial_z\sigma\partial_\zb\bar\sigma)
+{1\over e^2}\bar\beta\partial_z\beta\cr}}
We also need to make \eSW\ invariant under the $\Omega$ gauge
transformations. We'll see how to do that later. First, let's start looking at
some examples

\newsec{Examples}

\leftline{\bf Example 1: (2,2) Linear \sm\ on $\BC P^N$}

Choose $\Phi^i,\Lambda^i$, $i=1,\dots,N+1$ to all have charge $Q=1$.
After eliminating the auxiliary field $D$,
$$\eqalign{\CL&=({\rm kinetic\ terms\ for}\
\phi,\psi,\lambda,\alpha,\beta,\sigma)
+(\bar\alpha\bar\psi^i\phi^i-\alpha\psi^i\bar\phi^i)\cr
&\qquad+(\bar\beta\lambda^i\bar\phi^i-\beta\bar\lambda^i\phi^i)
+(\bar\lambda^i\psi^i\sigma-\lambda^i\bar\psi^i\bar\sigma)\cr
&\qquad+\sum_i|\phi^i|^2|\sigma|^2+{e^2\over 2}(\sum_i|\phi^i|^2-r)^2\cr
&\qquad +{1\over 2 e^2}f^2-{i\theta\over 2\pi} f}$$

We can analyse this theory semiclassically in the $r\gg0$ limit.
Supersymmetry requires that the scalar potential vanish, and hence that
\eqn\ecpn{\sum_i|\phi^i|^2=r,\qquad \sigma=0}
The space of solutions to \ecpn\ is a big sphere $S^{2N+1}$, but we still
must mod out by the action of the gauge transformations
$\phi\to\ex{i\theta}\phi$. So, after modding out, the $\phi$'s live on $\BC
P^N=S^{2N+1}/U(1)$. All of the degrees of freedom transverse to $\BC P^N$ have
masses of order $m^2\sim e^2r$. At energies well below this mass scale, we have
an effective {\it nonlinear} \sm\ with target space $\BC P^N$.

Well, that's what's happening with the bosons. Let us see what happens to the
fermions. Since $\phi$ has a  VEV, one linear combination of the $\psi$'s
gets a mass with $\alpha$. Which linear combination is it? Let
$\psi^i=\psi\phi^i$.
$$-\alpha\phi^i\bar\phi^i=-\alpha\psi|\phi|^2=-r\alpha\psi$$
so it is precisely the linear combination represented by $\psi$ which becomes
massive. The remaining $\psi^i$ transform as sections of the tangent bundle
to $\BC P^N$. Mathematically, the $\psi^i$ fit into the exact sequence
\eqn\etexact{0\to\CO{\buildrel\otimes\phi^i\over\longrightarrow}\CO(1)^{\oplus
N+1}\to T_{\BC P^N}\to 0}
Of course, exactly the same analysis holds for the left-moving fermions
$\lambda^i$.

\line{\hrulefill}

\leftline{\it Digression: Line bundles on $\BC P^N$}
Recall $\BC P^N=\BC^{N+1}/\sim$, where
$(z_0,z_1,\dots,z_N)\sim(\lambda z_0,\lambda z_1,\dots,\lambda z_N)$,
$\lambda\in\BC^*$. $\CO(-1)$, the ``tautological" line bundle has, as fiber
over the point $[z_0,z_1,\dots,z_N]\in\BC P^N$, the complex line through the
origin in $\BC^{N+1}$ which passes through $(z_0,z_1,\dots,z_N)$. Powers of
the tautological line bundle are denoted by $\CO(-n)=\CO(-1)^{\otimes n}$,
with negative powers denoting powers of the dual line bundle $\CO(1)$.
$\CO(1)$ is called the ``hyperplane bundle" because it has a global
holomorphic section which vanishes along a hyperplane, say
$\{z_0=0\}\subset\BC P^N$. The $\phi^i$ (and their superpartners $\psi^i$) in
the above example can be viewed as sections of $\CO(1)$.

\line{\hrulefill}

Very nice. Unfortunately for our intended application, $r$ is {\it not} a
renormalization group invariant in this model. Rather, there's a one-loop
log-divergent diagram \tfig\figone\ which contributes to the renormalization
of $r$. The $\beta$-function is proportional to the sum of the scalar charges
($\sum Q_i=N+1$ in this case), and the sign is such that $r(\mu)$ {\it
decreases} in the infrared. So even if we start out at large $r$, where the
theory is semiclassical, we don't stay there.

But this is exactly the sort of behaviour we expect. The $\BC P^N$ nonlinear
\sm\ also has a nonzero $\beta$-function and flows to strong coupling in the
infrared. It develops a mass gap, and the infrared theory is a $c=0$ CFT (a
topological field theory) \rgrassmannian.

\ifigure\figone{Log-divergent diagram leading to the
renormalization of $r$. Charged scalars run around the loop, and
the coefficient of the log-divergence is proportional to $\sum
Q_i$, the sum of the scalar charges.}{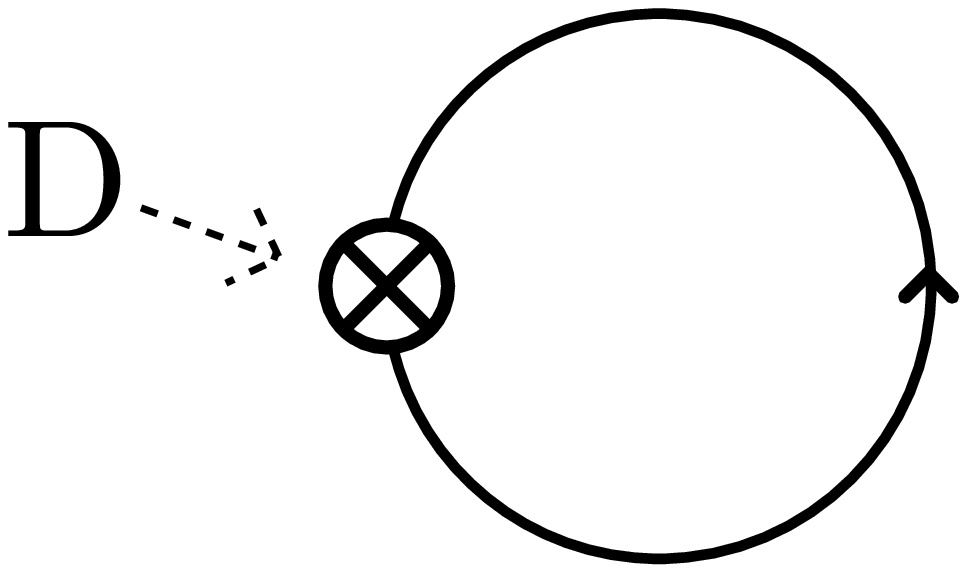}{1}

\leftline{\bf Example 2: Calabi-Yau hypersurface in $W\BP^4$}

As before, we consider $\Phi^i,\Lambda^i$, $i=1,\dots,5$, but now, instead of
taking them to all be of charge 1, we allow them to have (integer) charges
$w_i>0$. If we simply followed the analysis of example 1, we would obtain not
$\BC P^4$, but the {\it weighted} projective space $W\BP^4=\BC^5/\sim$, where
$$(z_1,\dots,z_5)\sim(\lambda^{w_1}z_1,\dots,\lambda^{w_5}z_5)$$

The sum of the scalar charges is still nonzero, so let us add a chiral scalar
superfield $P$, and fermionic superfield $\Gamma$ of charge $-d$, where
$d=\sum w_i$.
Under the $\Omega$-gauge transformation,
\eqn\eomegaGamma{\Gamma\to
\Gamma-2id\Omega P}
The action is as before, but now we can add a
gauge-invariant superpotential
\eqn\esmodtwo{S_\CW=\int d^2zd\theta^-\ \goth{m}(\Gamma W(\Phi)+\Lambda^i
P\pd{W}{\Phi^i})+ h.c.}
where $W(\Phi)$ is a weighted homogeneous polynomial of degree $d$ in the
$\Phi^i$. $\goth{m}$ is a parameter with dimensions of mass. We will, for the
most part, follow convention, and set it ``equal to one", but it is important
to remember that it is really there, setting the scale for certain of the
mass terms to be discussed below.

This is obviously invariant under the $\chi$-gauge transformations.
Invariance under $\Omega$-gauge transformations follows from
$$\sum_iw_i\Phi^i\pd{W}{\Phi^i}=d \ W(\Phi)$$

Adding the superpotential introduces new terms in the scalar potential from
integrating out the auxiliary fields in $\Lambda^i$ and $\Gamma$:
$$\CU={e^2\over 2}(\sum
w_i|\phi^i|^2-d|p|^2-r)^2
+|W|^2+|p|^2\left|\pd{W}{\phi^i}\right|^2+|\sigma|^2(\sum
w_i^2|\phi^i|^2+d^2|p|^2)$$
We also get some new Yukawa couplings:
$$\CL=\dots-(\gamma\psi^i+\lambda^i\pi)\pd{W}{\phi^i}-\lambda^i\psi^jp
{\partial^2W\over\partial\phi^i\partial\phi^j}+h.c.$$

We assume that $W$ is chosen to be transverse: $W=\pd{W}{\phi^i}=0\Rightarrow
\forall \phi^i=0$. The semiclassical analysis proceeds as before.

\leftline{\underbar{$r\gg0$}:}

Minimizing the scalar potential requires $\sum w_i|\phi^i|^2=r, p=\sigma=0$
and
$W(\phi)=0$. So after modding out by $U(1)$, the massless fields live on the
hypersurface $W(\phi)=0$ in $W\BP^4$.

The masses of the fields transverse to
the hypersurface are of the order $m^2\sim e^2r$ or $m^2\sim |\goth{m}|^2$,
depending  on whether they get a mass from the D-term or from the
superpotential. We will, for now, simply assume that these are of the same
order.

For the fermions, as before, one linear combination of $\psi^i$ gets a mass
with the gauge fermion. Another linear combination gets a mass from the
Yukawa coupling $-\gamma\psi^i\pd{W}{\phi^i}$. Mathematically, the massless
fermions form the cohomology of the sequence
$$0\to\CO{\buildrel f\over\longrightarrow}\bigoplus_i\CO(w_i){\buildrel
g\over\longrightarrow}\CO(d)\to 0$$
where $f(s)= (w_1\phi_1 s,\dots,w_5\phi_5 s)$ and $g(u_1,\dots,u_5)=\sum u_i
\pd{W}{\phi^i}$. this is precisely the sequence which defines the tangent
bundle of the  Calabi-Yau hypersurface: $T=\ker(g)/{\rm im}(f)$. So, as
expected,
the fermions transform as sections of the tangent bundle.

\leftline{\underbar{$r\ll0$}:}

Here too, we can find a supersymmetric vacuum by minimizing the scalar
potential. Set $|p|^2=|r|/d$, $\phi=\sigma=0$. $p$ and $\sigma$ are massive,
while the $\phi^i$ are massless. The left-moving fermion
$\gamma$ gets a mass with the gaugino $\bar\beta$ from
$\CL=\dots-d\bar\beta\gamma\bar p+\dots$. The right-moving fermion $\pi$ gets
a mass with the gaugino $\alpha$ from $\CL=\dots +d\alpha\pi\bar p+\dots$.

The low energy theory is described by the superpotential
$$\int d^2zd\theta^-\ {\rm const}\ \Lambda^i\pd{W}{\Phi^i}+ h.c.$$
This should be recognizable as the superpotential for a (2,2) \LG\ theory.
Actually \phases,
it is a \LG\ {\it orbifold}.  Since $p$ has charge $-d$, its VEV
doesn't completely break the gauge symmetry. Gauge transformations by
$d^{th}$ roots of unity are still unbroken and so we should still mod out
our \LG\ theory by this unbroken $\BZ_d$ group. Projecting onto the
$\BZ_d$-invariant states requires, for modular invariance, that we introduce
twisted sectors, with boundary conditions twisted by
$\BZ_d$\foot{Alternatively,
on a higher genus Riemann surface, we sum over sectors where the boundary
conditions on the fields around each cycle
are twisted by $\BZ_d$ gauge transformations.}

\leftline{\bf Example 3: Deformations of (2,2) theories}

The superpotential \esmodtwo\ was not the most general one compatible with
the $\Omega$ gauge transformations. More generally,
\eqn\esmodthree{S_\CW=\int d^2zd\theta^-\ (\Gamma W(\Phi)+\Lambda^i
PF_i(\Phi))+ h.c.}
where
\eqn\emodthreecond{\sum w_i\phi^iF_i(\phi)=d\ W(\phi)}
is invariant under $\Omega$-gauge transformations. Generically, this breaks
(2,2) supersymmetry down to (0,2).
For example, if $W(\phi)$ is a quintic polynomial in $\BC P^4$, there is a
224-dimensional space of polynomials $F_i$ satisfying \emodthreecond.

For $r\gg0$, we see that the right-moving fermions $\psi^i$ which remain
massless again transform as sections of $T$, but the massless left-moving
fermions $\lambda^i$ transform as sections of $V$ (a holomorphic deformation
of $T$), which is the cohomology of the sequence
$$0\to\CO{\buildrel \otimes
w_i\phi^i\over\longrightarrow}\bigoplus_i\CO(w_i){\buildrel
\otimes F_i(\phi)\over\longrightarrow}\CO(d)\to 0$$

These are, to be sure,
(0,2) theories, {\it but} the rank of $V$ remains $r=3$, $E_6$ is unbroken and
(one can show)  the number of {\bf 27}s and $\overline{\bf 27}$s remains
unchanged.

\leftline{\bf Example 4: Dispensing with the $\Sigma$ multiplet}
We needed the $\Sigma$ multiplet and the accompanying $\Omega$ gauge
transformations in order to describe (2,2) supersymmetric theories. If we're
really interested in (0,2) theories, why not dispense with them and consider
\eqn\esmodfour{S_\CW=\int d^2zd\theta^-\ (\Gamma W(\Phi)+\Lambda^i
PF_i(\Phi))+ h.c.}
where, now, freed from the constraint of $\Omega$ gauge-invariance, the
$F_i(\phi)$ are arbitrary polynomials of the appropriate degree.

Recall that previously there were two mass terms for the left-moving fermions
$$\CL=\dots +w_i\bar\beta\lambda^i\bar\phi^i-\lambda^i\pi F_i(\phi) +h.c.$$
Now there's only one (since $\bar\beta$ is absent from the theory). So $V$ is
defined by
\eqn\Vseq{0\to V\to\oplus_i\CO(w_i){\buildrel \otimes
F_i(\phi)\over\longrightarrow}\CO(d)\to 0}
and now has rank $r=4$.

It turns out that this theory is ill-behaved; we'll see why later.

\leftline{\bf Example 5: Arbitrary charges}

Since the $\Lambda$s are supposed to be unrelated to the $\Phi$s, it is silly
to give them the same label and to assume that they have the same gauge
charges. So let
\eqn\esmodfive{S_\CW=\int d^2zd\theta^-\ (\Gamma W(\Phi)+\Lambda^a
PF_a(\Phi))+ h.c.}
where now we let the charges of the fields be given in the table below.

\hbox{\vtop{\hsize=2.25in
\def\tablerule{\omit&
\multispan{4}{\tabskip=0pt\hrulefill}&\cr}
\def\tablepad{\omit&
height3pt&&&&\cr}
$$\vbox{\offinterlineskip\tabskip=0pt\halign{
\hskip.5in
\strut$#$\quad&
\vrule#&\quad\hfil $#$ \hfil\quad &\vrule #&\quad
\hfil $#$ \quad&\vrule #\cr
&\omit&\hbox{Field}&\omit&Q&\omit\cr
\tablerule\tablepad
&& \Phi^{i} &&w_{i}&\cr
\tablepad\tablerule\tablepad
&&P&&-m&\cr
\tablepad\tablerule\tablepad
&&\Lambda^{a}&&n_{a}&\cr
\tablepad\tablerule\tablepad
&&\Gamma&&-d&\cr
\tablepad\tablerule
\noalign{\bigskip}
\noalign{\narrower\noindent{\bf Table 2:}
$U(1)$ charges of the (bosonic and fermionic) chiral superfields.}
 }}$$}
\vtop{\advance\hsize by -2.25in
We still require
\eqn\linconds{\sum_i w_i=d,\qquad\sum_a n_a=m}
which in the semiclassical
$r\gg0$ analysis are equivalent to the conditions $c_1(T)=c_1(V)=0$. But
now, since we are dealing with a theory with chiral fermions coupled to a
gauge field, we need to ensure that there is no gauge anomaly. This is a
quadratic condition on the gauge charges:}}

\epsfxsize=1in
$$ \hskip-.125in\vcenter{\hbox to 1.125in{\epsfbox{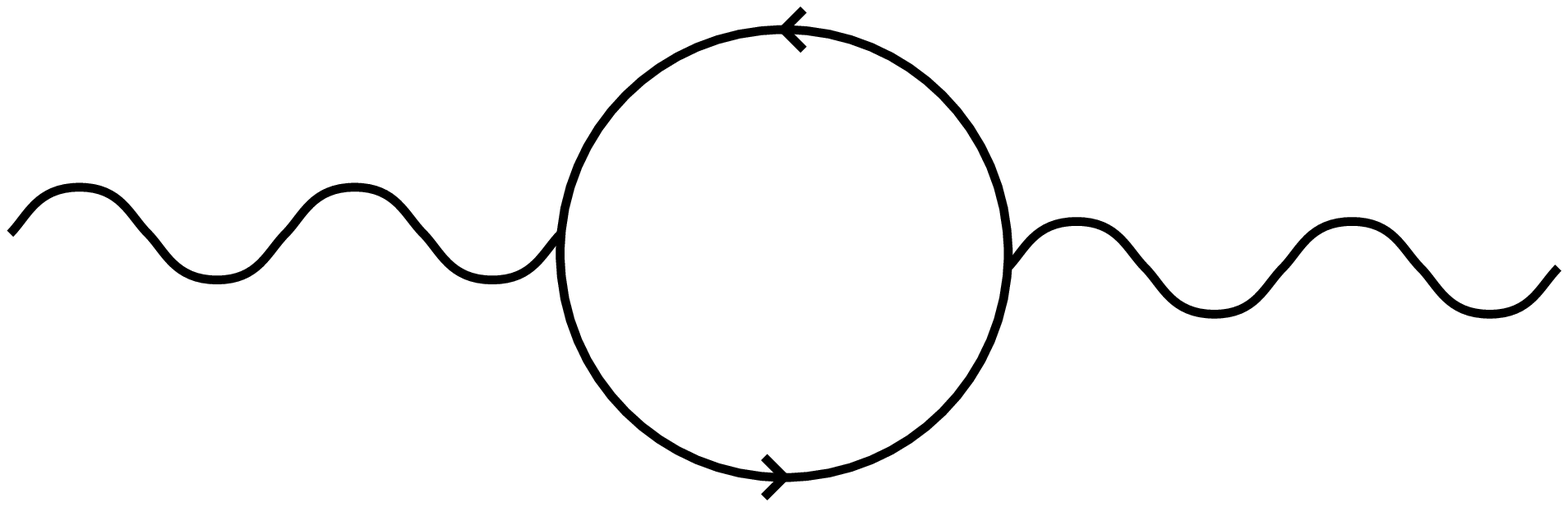}}}\propto\underbrace{\sum
n_a^2+d^2}_{\rm left-movers} -\underbrace{\sum w_i^2-m^2}_{\rm right-movers}\equiv0$$

Rearranging this,
$$\half(-m^2+\sum n_a^2)=\half(-d^2+\sum w_i^2)$$
which, in the $r\gg0$ \cy\ phase is simply the condition $c_2(V)=c_2(T)$!
This is a general principle which holds in all of the (0,2) linear \sm s.
The condition for the cancellation to worldsheet
gauge anomalies translates into the $c_2(V)=c_2(T)$ condition for the vanishing
of \sm\ anomalies.

The exact sequence defining the bundle $V$ is now
\eqn\Vseqtwo{0\to V\to\oplus_i\CO(n_a){\buildrel \otimes
F_a(\phi)\over\longrightarrow}\CO(d)\to 0}
instead of \Vseq. And now, most importantly, the rank of $V$ and the number
of generations independent of those of the tangent bundle.

Perhaps more surprising is the \LG\ phase for $r\ll0$. There, $p$ has a VEV,
and since it has charge $-m$, this breaks $U(1)\to \BZ_m$. But since $m\neq d$
in general, the orbifold group is {\it different} from that of the
corresponding
(2,2) model. This means, in particular, that the \Ka\ moduli space is
topologically different from that of the (2,2) model.  At large positive $r$,
they are obviously the same, as they both describe the space of complexified
\Ka\ forms for a weakly coupled \cy\ \sm. But globally, the CFT ``knows" the
difference between the ``\Ka " degrees of freedom of the (2,2) and the (0,2)
models.

\leftline{\bf Example 6: Return of the $\Sigma$ field}

We can also construct models which include the $\Sigma$ multiplet. let us
simply {\it postulate} a transformation law under the $\Omega$-gauge
transformations of the form
\eqn\eLarbOm{\Lambda^a\to\Lambda^a+2i\Omega E^a(\Phi)}
where $E^a(\phi)$ is a polynomial of weighted degree $n_a$. Then the
action
$$\eqalign{S_\Lambda+S_\Sigma=&\int d^2zd^2\theta\
\ex{2n_aV}(\bar\Lambda^a\Lambda^a +
|E^a(\Phi)|^2\bar\Sigma\Sigma-E^a(\Phi)\bar\Lambda^a\Sigma
-\overline{E^a(\Phi)}\Lambda^a\bar\Sigma)\cr
&\qquad+{1\over2e^2}(-\bar D_-\bar\Sigma\partial_z\bar D_+\Sigma
+\partial_z\bar D_-\bar \Sigma\bar D_+\Sigma)\cr}$$
is invariant under the $\Omega$-gauge symmetry. The superpotential
\esmodthree\ is invariant, provided
$$E^a(\phi)F_a(\phi)= d\ W(\phi)$$
$V$ is now the cohomology of the sequence
\eqn\eVseqthree{0\to\CO{\buildrel \otimes E^a(\phi)\over \longrightarrow}
\bigoplus_a\CO(n_a){\buildrel \otimes F_a(\phi)\over\longrightarrow}\CO(m)\to
0}
This is not so obvious as before, when we were construction the tangent
bundle $T$.  It is clear that one linear combination of the $\lambda^a$ get a
mass from the $-\lambda^a\pi F_a(\phi)$ term coming from the superpotential,
and that this corresponds to their being in the kernel of the map to $\CO(m)$
in \eVseqthree. To see that the other linear combination of the $\lambda^a$
drop out as required, it is easiest to not work in W.Z.~gauge, but instead to
use the $\Omega$-gauge symmetry \eLarbOm\ to gauge them away. The remaining
$\lambda^a$ are clearly in the quotient by the image of $\CO$ in \eVseqthree.

\leftline{\bf Example 7: Complete intersection \cy\ manifolds}

Our previous examples have, in the \cy\ phase, corresponded to (0,2) models
defined on hypersurfaces in $W\BP^4$. There is clearly no need to restrict
ourselves to hypersurfaces. Complete intersection \cy's are just as easy
to describe.

Instead of a single $\Gamma$, let there be several $\Gamma^\alpha$, and let the
superpotential be
\eqn\esmodfive{S_\CW=\int d^2zd\theta^-\ (\Gamma^\alpha W_\alpha(\Phi) +
\Lambda^a P F_a(\Phi))}

As before, we require
\eqn\linconds{\eqalign{
-\sum d_\alpha+\sum_i w_i=0\quad &\Leftrightarrow\quad c_1(T)=0\cr
-m+\sum_a n_a=0\quad &\Leftrightarrow\quad c_1(V)=0\cr
\sum d_\alpha^2-\sum w_i^2=m^2-\sum n_a^2\quad &\Leftrightarrow\quad
c_2(T)=c_2(V)\cr}}

\hbox{\vtop{\hsize=2.25in
\def\tablerule{\omit&
\multispan{4}{\tabskip=0pt\hrulefill}&\cr}
\def\tablepad{\omit&
height3pt&&&&\cr}
$$\vbox{\offinterlineskip\tabskip=0pt\halign{
\hskip.5in
\strut$#$\quad&
\vrule#&\quad\hfil $#$ \hfil\quad &\vrule #&\quad
\hfil $#$ \quad&\vrule #\cr
&\omit&\hbox{Field}&\omit&Q&\omit\cr
\tablerule\tablepad
&& \Phi^{i} &&w_{i}&\cr
\tablepad\tablerule\tablepad
&&P&&-m&\cr
\tablepad\tablerule\tablepad
&&\Lambda^{a}&&n_{a}&\cr
\tablepad\tablerule\tablepad
&&\Gamma^\alpha&&-d_\alpha&\cr
\tablepad\tablerule
\noalign{\bigskip}
\noalign{\narrower\noindent{\bf Table 3:}
$U(1)$ charges of the (bosonic and fermionic) chiral superfields.}
 }}$$}
\vtop{\advance\hsize by -2.25in
For simplicity, we assume that we are working in a model without the $\Sigma$
multiplet and the associated $\Omega$-gauge symmetry.
In the semiclassical $r\gg0$ analysis, this superpotential leads to a (0,2)
model where the bundle $V$ is defined by the sequence \Vseqtwo, living on the
complete intersection $W_\alpha(\phi)=0$.}}

\leftline{\bf Spectators:}

In the previous three examples, one generally finds that the sum of the scalar
charges is nonzero. In a (2,2) model, this would be a fatal flaw. In the
present context, we can always fix this (so that there is no perturbative
renormalization of $t$) by adding a pair of chiral superfields $S,\Xi$.

\hbox{\vtop{\hsize=2.25in
\def\tablerule{\omit&
\multispan{4}{\tabskip=0pt\hrulefill}&\cr}
\def\tablepad{\omit&
height3pt&&&&\cr}
$$\vbox{\offinterlineskip\tabskip=0pt\halign{
\hskip.25in
\strut$#$\quad&
\vrule#&\quad\hfil $#$ \hfil\quad &\vrule #&\quad
\hfil $#$ \hfil\quad&\vrule #\cr
&\omit&\hbox{Field}&\omit&Q&\omit\cr
\tablerule\tablepad
&& S&&m-\sum d_\alpha&\cr
\tablepad\tablerule\tablepad
&&\Xi&&-m+\sum d_\alpha&\cr
\tablepad\tablerule
\noalign{\bigskip}
\noalign{\narrower\noindent{\bf Table 4:}
$U(1)$ charges of the ``spectators". $S$ is a bosonic chiral superfield,
and $\Xi$ a fermionic chiral superfield.}
 }}$$}
\vtop{\advance\hsize by -2.25in
To the superpotential, we add a term
\eqn\eswpert{S_\CW=\int d^2zd\theta^-\ \dots +m_s\Xi S}
Examining the scalar potential, $\CU$, we see that $s=0$, and all the
fluctuations of $S$ and $\Xi$ are massive. So, naively, they do not affect
the low-energy physics.
}}

This is a little too slick. In the \LG\ phase, we
see that $S,\Xi$ are charged under the unbroken $Z_m$ symmetry, so their
boundary conditions are twisted in the twisted sectors of the orbifold.
Nonetheless, since they appear only quadratically in the superpotential,
when we compute the cohomology of the $\bar Q_+$ operator in \S7, there is
a choice of representatives in which the $S,\Xi$ oscillators are not excited.
In particular, this means that they do not appear in the massless states of
the string theory. They also make no {\it net} contribution to the ground state
energies or $U(1)$ charges of the twisted sectors.

This seems to pose a small paradox. What if we decided to make $m_s$ very
large? Surely, below the scale of $m_s$, we should be able to describe the
physics in terms of an effective theory with $\Xi,S$ absent.  Aren't we then
back in the situation where $t={\theta\over 2\pi}+ir$ is scale-dependent (as in
{\bf Example 1})?

Actually, we might have asked a similar question already in the (2,2) theory
we discussed as example 1. There we had two, {\it a priori independent}, mass
scales $e$ and $\goth{m}$ (where
$\goth{m}$ is the parameter introduced in \esmodtwo). If we take these to be
disparate, we should work with an effective theory in the intervening regime,
and in that effective theory, we expect that $t$ will run. Put another way, we
might expect the low-energy \Ka\ class to depend on the dimensionless ratio
$\goth{m}/e$.

Is this, in fact, the case?  The answer, of course, is no, and the reason is as
follows. The Wilsonian coupling, $t_{LE}$, being the coefficient of a
term in the superpotential, is an analytic function of $\goth{m}$ \SeiNat. So,
though we're really interested in the dependence on the magnitude of
$\goth{m}$,
we can equally well inquire about its dependence on the {\it phase} of
$\goth{m}$. But the phase of $\goth{m}$ can be rotated away by a common
rotation
of the superfields $\Gamma$ and $P$. This symmetry is nonanomalous, so the
phase
of
$\goth{m}$ is unphysical, and $t_{LE}$ must be independent of it (and hence of
$\goth{m}$ itself).

In the (0,2) theories under consideration, there is no nonanomalous symmetry
which allows us to simultaneously remove both the phase of $\goth{m}$ and of
$m_S$. The ratio, $\goth{m}/m_S$ is physical. More precisely, the combination
\eqn\tphysdef{t_{phys}= t-{i\over 2\pi}Q_s\ln (\goth{m}/m_s)}
where $Q_s=m-\sum d_\alpha$ is the charge of the spectator scalar $S$, is
physical. A change in the phase of $m_s$ can be compensated by a shift in the
$\theta$-angle (the real part of $t$). Moreover, we can choose a basis in which
this is the only anomalous $U(1)$.

If we choose to work with an effective theory in the range
$|\goth{m}|<\mu<|m_S|$, then, of course, $t(\mu)$ runs:
\eqn\trun{t(\mu)={i\over 2\pi}Q_S\ln (\Lambda/\mu)}
with $\Lambda=m_S \ex{-2\pi it/Q_S}=\goth{m}\ex{-2\pi it_{phys}/Q_S}$.
$t$ is no longer RG-invariant, but $t_{phys}$ is.

Note what's going on here. The dependence of the real part of $t$ on the
parameters in the superpotential is governed by the chiral anomaly (which,
by the way, receives contributions only at one loop).  But, by holomorphy, this
constrains the dependence of the
imaginary part of
$t$ as well\foot{Unlike in N=1 supersymmetric theories in four dimensions,
where the analogous statement is true only in a very special renormalization
scheme \ShifVain, here (because the theory is all but superrenormalizable)
it is true without any delicacies in the argument. For a recent discussion of
the scheme-dependence of the four dimensional result, see \DineHolo.}. Thus
$\trun$ is an exact expression.

So $\dots$ the upshot is that there is {\it one} physical, RG-invariant
parameter
\tphysdef\ which parametrizes the low energy physics, rather than the two
``naive" parameters $t$ and $\goth{m}/m_S$. Of course, if we are smart, we
simply use our freedom to set $m_S=\goth{m}$, in which case $t$ and $t_{phys}$
coincide. In that case, the ``spectators" should be though of as being
on exactly
the same footing as the other massive particles in the linear \sm.

Now, $t_{phys}$ is the RG-invariant parameter in the linear \sm\ which
parametrizes the ``\Ka\ moduli space". However, if we wish to discuss the low
energy physics in terms of the effective nonlinear \sm, we need to perform a
matching between $t_{phys}$ and the complexified \Ka\ parameter of the
nonlinear \sm. At tree level, they are simply equal, and indeed this equality
holds to all orders in perturbation theory. If we write $z=\ex{2\pi i
t_{phys}}$, and $q=\ex{2\pi i \CJ}$, then we have $q=z$. However,
nonperturbatively, this relation is modified to
$$q=z(1+\sum_{n=1}^\infty a_n z^n)$$
This is easily recognized as an instanton correction to the matching
condition (a holomorphic $n$-instanton effect goes like $z^n$). Its origin
is simply the fact that the linear \sm\ possesses instanton solutions (dubbed
``pointlike instantons" in \phases) which are not present in the nonlinear
\sm. To perform the matching, we need to integrate out the pointlike
instantons, which introduces a nontrivial matching condition. Note that this
is not a feature peculiar to (0,2) theories. It occurs as well in (2,2)
theories, where it is called the relation between the algebraic and \sm\
coordinates \MonomialDivisor\ (For a recent discussion from the point of view
of linear \sm s, see \MorrisonPlesser).

\leftline{\bf Onward!}

There are numerous variations on the constructions described here.
We can have more $U(1)$ gauge groups, and thereby construct complete
intersection \cy s in more general toric varieties. We can also generalize the
construction of the gauge bundle $V$. By mixing and matching the various
constructions, one can produce a wide variety of different models (see
\refs{\phases,\DK}). This may still only be scratching the surface of the
space of (0,2) models.

Having discussed the tools for writing down (0,2) models, we will now turn to
computing some of their properties. The key, as discussed in \S3, is to
consider RG-invariant quantities which can, reliably, be calculated in the
linear \sm.

A rich class of interesting things to calculate can be found by considering
the ``twisted" model, or, equivalently, to compute on the cylinder with
periodic boundary conditions on the right-movers. This is the sector of the
String Hilbert space containing the spacetime fermions. In this sector,
(0,2) supersymmetry is unbroken, and the supercharges, $\bar Q^\pm$, close
into the generator of boosts along the lightcone,
$$\{\bar Q^+,\bar Q^-\}=\bar L_0$$
If we are interested in those states with $\bar L_0=0$, which includes all
of the massless spacetime fermions, we can represent these as the cohomology
of the $\bar Q^+$ operator. We will see that the $\bar Q^+$-cohomology is
eminently computable, so that, in particular, we learn about the spectrum of
massless spacetime fermions and (by spacetime supersymmetry) about the full
massless spectrum of the string theory.

 Similarly, we can compute the matrix elements of $\bar Q^+$-invariant
operators between these fermion states, which, in particular, allows us to
calculate the spacetime superpotential.

\newsec{\LG\ }

As we saw, the semiclassical analysis becomes exact for $r\to\pm\infty$.
$r\to+\infty$ is a weakly-coupled (0,2) nonlinear \sm. Much of what we
currently
can say about such \sm s has been understood for many years \twozero. Instanton
corrections are also suppressed in the
$r\to -\infty$ limit. Here, too, one can make some definite statements, at
least in those cases where the conformal field theory that one obtains in the
$r\to-\infty$ limit is understood. For the examples discussed in the previous
section, the $r\to-\infty$ limit corresponds to what we might call a (0,2)
\LG\ orbifold. In this case, we actually have a fair handle on the conformal
field theory, and can actually make some definite statements.

To describe the low energy theory, we need, in particular, two {\it unbroken}
nonanomalous $U(1)$ symmetries. One will be the $U(1)_R\subset(\hbox{
right-moving $N=2$} )$. The other will become the left-moving $U(1)$ which we
introduced in \S2. Denoting the charges under the two $U(1)$, respectively, as
$\qr,\ql$, it is clear that in order for these to be symmetries of the
superpotential
$$S_\CW=\int d^2 zd\theta^-\ \Gamma^\alpha W_\alpha(\Phi)
+\Lambda^aF_a(\Phi)+\Xi S$$
 the charges of the $\Phi_i$ must be proportional to
the respective gauge charges, $q_i=w_i/a,\ \bar q_i= w_i/b$. Since the left
$U(1)$ is an honest symmetry, and not an R-symmetry, we can, without loss of
generality, rescale the left $U(1)$ charge s.t. $q_i=\bar q_i$, or $a=b$. We
will see later that this corresponds to the ``standard" normalization of the
corresponding current, $J$. The charges of the fermions are now determined, up
to
this unknown constant $b$.

We need to check, first of all, that these $U(1)$'s are nonanomalous under the
gauge symmetry. The anomaly, of course, is given by a one-loop diagram with one
insertion of the current, one external gauge field, and fermions running around
the loop:
$$\eqalign{J \hskip-.125in\vcenter{\epsfxsize=.75in\hbox to .875in{\epsfbox{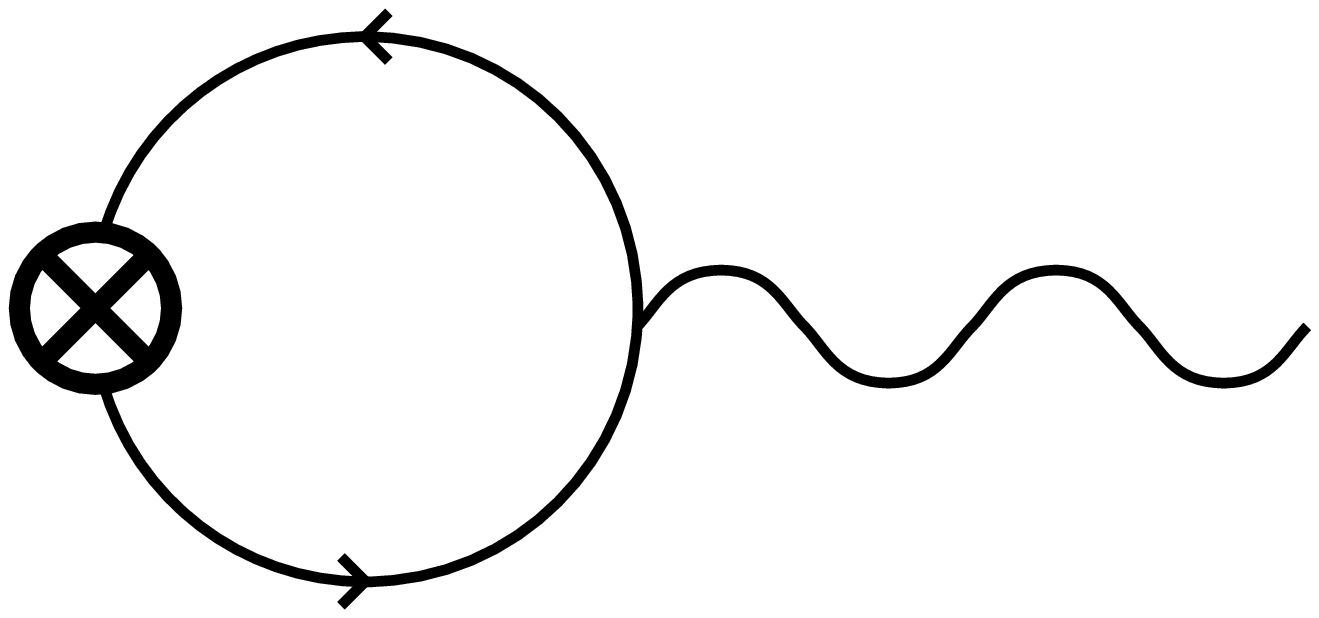}}}&\propto
{1\over b}[\sum (n_a-m)n_a +\sum d_\alpha^2 -\sum w_i^2] =0\cr
\bar J \hskip-.125in\vcenter{\epsfxsize=.75in\hbox to .875in{\epsfbox{anom2.eps}}}&\propto {1\over
b}[\sum (n_a-m+b)n_a +\sum (d_\alpha-b)d_\alpha -\sum (w_i-b)w_i] -m =0\cr
}$$
and these symmetries are nonanomalous, precisely when the conditions \linconds\
on the gauge charges are satisfied. Note that, since $\bar J$ is an
R-symmetry, even though $p$ is neutral, its fermi superpartner, $\pi$, is not,
and contributes to the anomaly. Note also that the ``spectators" make no net
contribution to the anomalies.

 We also need to require that
$J$ and
$\bar J$ are pure left- and right-moving currents in the infrared, which means
that the mixed anomaly vanishes:
$$\eqalign{J \hskip-.125in\vcenter{\epsfxsize=.5in\hbox to.625in{\epsfbox{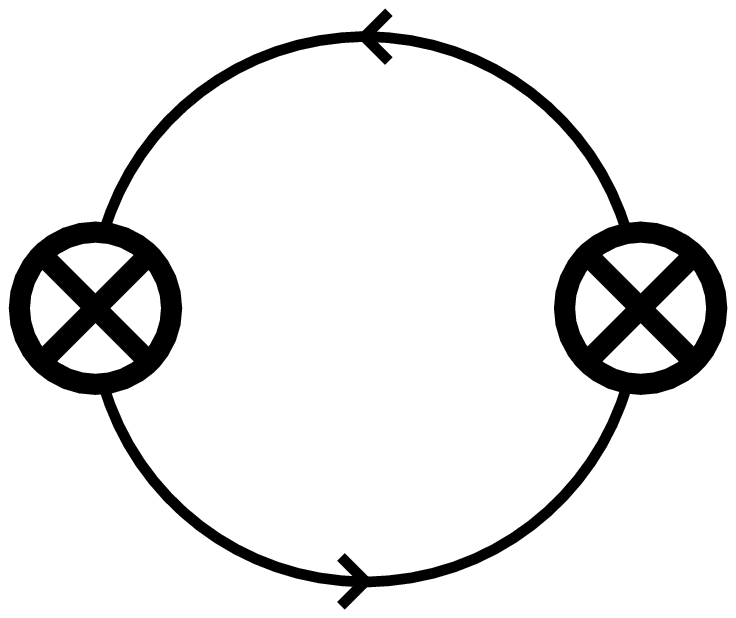}}}\bar J\propto&
{1\over b^2} [\sum (n_a-m)(n_a-m+b)+\sum d_\alpha(d_\alpha-b) -\sum
w_i(w_i-b)]\cr=&-{rm(m-b)\over b^2}}$$
For this to vanish, we must have $b=m$. So the charges of the fields are listed
in table 5.

\hbox{\vtop{\hsize=2.625in
\def\tablerule{\omit&
\multispan{6}{\tabskip=0pt\hrulefill}&\cr}
\def\tablepad{\omit&
height3pt&&&&&&\cr}
$$\vbox{\offinterlineskip\tabskip=0pt\halign{
\hskip-.25in
\strut$#$\ \ &
\vrule#&\ \ \hfil $#$ \hfil\ \ &\vrule #&\ \
\hfil $#$ \hfil\ \ &\vrule #&\ \
\hfil $#$ \hfil\ \ &\vrule #\cr
&\omit&\hbox{Field}&\omit&\ql&\omit&\qr&\omit\cr
\tablerule\tablepad
&&\Phi_i&&q_i={w_i\over m}&&\bar q_i={w_i\over m}&\cr
\tablepad\tablerule\tablepad
&&\Lambda^a&&q_a={n_a\over m}-1&&\bar q_a={n_a\over m}&\cr
\tablepad\tablerule\tablepad
&&\Gamma^\alpha&&q_\alpha=-{d_\alpha\over m}-1&&\bar q_\alpha=
1-{d_\alpha\over m}&\cr
\tablepad\tablerule\tablepad
&&P&&0&&0&\cr
\tablepad\tablerule\tablepad
&&S&&{Q_S\over m}&&{Q_S\over m}&\cr
\tablepad\tablerule\tablepad
&&\Xi&&-{Q_S\over m}&&1-{Q_S\over m}&\cr
\tablepad\tablerule
\noalign{\bigskip}
\noalign{\narrower\noindent{\bf Table 5:}
Left-moving $U(1)$ and right-moving $U(1)_R$ charges of the fields.}
 }}$$}
\vtop{\advance\hsize by -2.625in
We have cheated a bit. The anomaly considerations do not determine the charges
of
$\Xi,S$. But recall that, because of the unbroken $\BZ_m$
discrete gauge symmetry of
the model, we are really describing a \LG\ orbifold. The generator of $\BZ_m$
is simply $\ex{-2\pi i \ql}$. Since we know how $\Xi,S$ are supposed to
transform under  $\BZ_m$, this fixes their charges, $\ql$, modulo 1.

In fact, for our
purposes, it is not useful to separate the $\BZ_m$ orbifolding from the $\BZ_2$
orbifolding which implements the GSO projection. Together, they form a
$\BZ_{2m}$ group generated by
}}

\eqn\gsoLGdef{
g=\ex{-i\pi \ql}(-1)^{F_{\lambda^I}}
}
Comparing with \gsodef, we see that this is, indeed, the standard normalization
of the left-$U(1)$ charge which we defined in \S2.

Now we can calculate the central charge of the infrared $N=2$ superconformal
algebra. Recall that one of the OPEs in the $N=2$ superconformal
algebra is
$$\bar J(\zb)\bar J(\wb)={\bar c/3\over(\zb-\wb)^2}$$
So, by calculating this OPE, we get a direct measurement of the central charge.
But this is, as we have seen, computable from the one loop anomaly diagram in
the linear \sm, $\bar J \hskip-.125in\vcenter{\epsfxsize=.25in\hbox
to.375in{\epsfbox{anom3.eps}}}\bar J$.
$${\bar c\over 3}=\sum (\bar q_i-1)^2-\sum \bar q_a^2-\sum\bar q_\alpha^2=3$$
So, indeed, we have $\bar c=9$, as expected!

Similarly, the $J\cdot J$ anomaly computes the level of the left-moving $U(1)$
current algebra:
$$r=\sum q_\alpha^2+\sum q_a^2 -\sum q_i^2$$
In fact, we can make an even stronger statement. The operators
\eqn\offshell{\eqalign{
T'&=-\sum_i\left(\partial\phi_i\partial\bar\phi_i+{q_i\over2}\partial(\phi_i
\partial\bar\phi_i)\right)+\sum_a\left(\lambda_a\partial\bar\lambda_a
+{q_a\over 2}\partial(\lambda_a\bar\lambda_a)
\right)\cr
&\phantom{=-\sum_i\left(\partial\phi_i\partial\bar\phi_i
+{q_i\over2}\partial(\phi_i
\partial\bar\phi_i)\right)}
+\sum_\alpha\left(\gamma_\alpha\partial\bar\gamma_\alpha+{q_\alpha\over
2}\partial(\gamma_\alpha\bar\gamma_\alpha)
\right)
\cr
J'&=-\sum_i q_i\phi_i\partial\bar\phi_i-\sum_a q_a\lambda_a\bar\lambda_a
-\sum_\alpha q_\alpha \gamma_\alpha\bar\gamma_\alpha
\cr}}
commute with the $\bar Q^+$ operator, and generate a $\widehat{U(1)}\sdp
virasoro$ algebra on the $\bar Q^+$ cohomology, with $\widehat{U(1)}$ level
$r$ and virasoro central charge $c=6+r$. To see this, note that rescaling the
superpotential is a
$\bar Q^+$-trivial operation. Thus, while working on the level of the $\bar
Q^+$ cohomology, we can use free-fields to evaluate the OPEs of $T',J'$.

Note, too, that $J'_0$ differs from the previously-defined $U(1)$ charge $\ql$
by $\bar Q^+$-trivial terms, as does $L'_0$ from the $L_0$ derived
canonically from the Lagrangian. Thus, for instance, since we will be
working on the
$\bar Q^+$-cohomology, we can continue to use $\ql$ to label the $U(1)$ charge
of physical states.

Not only does this ``off-shell" virasoro algebra exist at the \LG\ point,
Silverstein and Witten have shown that one can construct the corresponding
operators in the full linear \sm\ \eva. So the fact that one gets the correct
infrared virasoro central charge and $\widehat{U(1)}$ level is a property of
the linear \sm\ for arbitrary $r$, not just for $r\to -\infty$. Indeed this
can be used as the basis for an argument that the  (0,2) linear \sm s do
indeed give rise to (0,2) SCFTs in the infrared, and that deforming $t$
really is an exactly-marginal deformation of the SCFT \DKtwo.

\leftline{\bf GSO projection:}

We have already said that to effect the left-moving GSO projection of this
theory (while simultaneously implementing the discrete $\BZ_m$ gauge
symmetry), we orbifold by the $\BZ_{2m}$ group generated by
$$g=\ex{-i\pi \ql}(-1)^{F_{\lambda^I}}$$
(All the charges in the theory are multiples of $1/m$, so $g^{2m}=1$.)
As usual, we must include $2m-1$ twisted sectors, where the boundary conditions
on the fields are twisted by powers of $g$. So the sectors of the theory are
labeled  by $k$, $k=0,\dots,2m-1$. $k$ {\it even} corresponds to the
left-moving Ramond sector, $k$ {\it odd} corresponds to the left-moving
Neveu-Schwarz sector. The right-mover will, for us, always be in  the Ramond
sector, so we are discussing states which are spacetime fermions. And the
$\bar Q^+$ cohomology computes those states which are physical and have $\bar
L_0=0$.

\leftline{\bf Modular Invariance}

The usual level-matching conditions for orbifolds \levelmatching, which are
certainly satisfied by our constructions are not {\it obviously} sufficient
here to assure the consistency of the theory, {\it unlike} the case of the
usual toroidal orbifolds.

There's a general principle in string theory that there is a direct
correspondence between worldsheet and spacetime anomalies. We will look for a
spacetime anomaly in the $\BZ_{2m}$ quantum symmetry \VafaQ\ of the \LG\
orbifold. This is a discrete R-symmetry in spacetime. As such, it may not be
preserved in a nontrivial gauge or gravitational background. That is, it may
suffer from an anomaly.  But the quantum symmetry (which simply says that
sector number must be conserved {\it modulo $2m$}) in any correlation
function) {\it must} be a symmetry if the GSO-projected theory is to have a
sensible interpretation.

Naively, the generator of the quantum symmetry is $\gamma=\ex{2\pi i k/2m}$,
where
$k=0,\dots,2m-1$ is the sector number. However, it proves more convenient to
compose this with a gauge transformation which lies in the $U(1)$ subgroup of
the spacetime gauge group generated by $\ql$:
\eqn\quantsymdef{\gamma=\ex{2\pi i(k r-2\ql)/2m r}}
With this choice, irreducible representations of the spacetime gauge group
transform homogeneously under the quantum symmetry.

One can compute the anomaly in this discrete R-symmetry by embedding it in a
continuous $U(1)$ R-symmetry (with generator $(kr-2\ql)$) and computing the
standard triangle diagram, with one insertion of this current and two
external gauge bosons or gravitons. Of course, since we are really only
interested in assuring that the discrete subgroup is nonanomalous, we need
only require the anomaly coefficient to vanish $mod\ 2mr$, rather than
actually vanish.
\eqn\threeanoms{\eqalign{
{\rm  A}_1&=\hskip-.125in\vcenter{\epsfxsize=1in\hbox to1.125in{\epsfbox{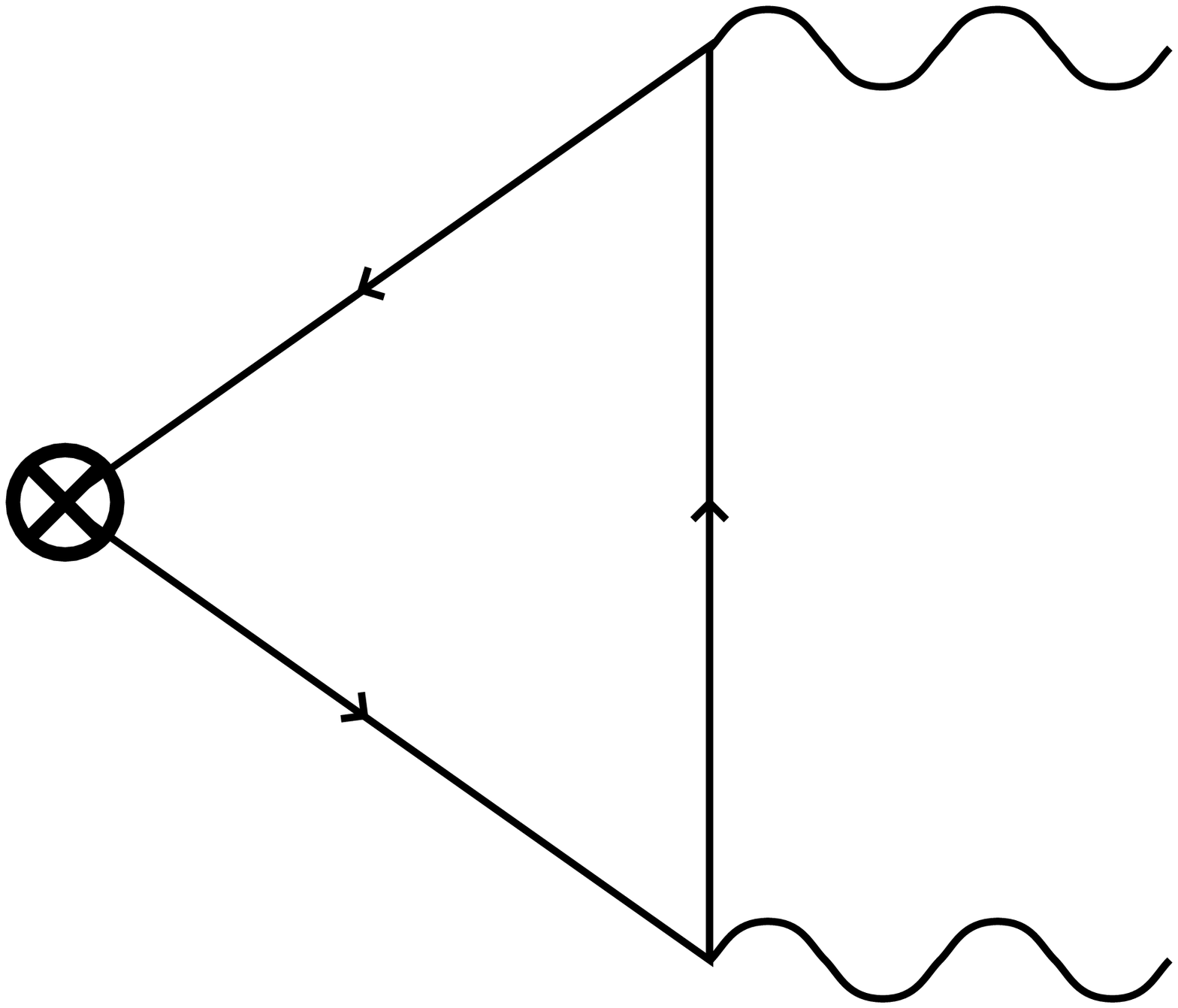}}}
G\hbox{-gauge bosons\hskip .5in} mod\ 2mr\cr
{\rm A}_2&=\hskip-.125in\vcenter{\epsfxsize=1in\hbox to1.125in{\epsfbox{triangleanom.eps}}}
E_8\hbox{-gauge bosons\hskip .45in} mod\ 2mr\cr
{\rm A}_3&=\hskip-.125in\vcenter{\epsfxsize=1in\hbox to1.125in{\epsfbox{triangleanom.eps}}}
\hbox{gravitons\hskip
.9in} mod\ 2mr\cr
 }}
 where $G$ is the ``observable" gauge group ($E_6$, $SO(10)$, or $SU(5)$,
for $r=3,4,5$, respectively).

Each of these can be expressed in terms of a
trace over the right-moving R-sector. For instance, $${\rm A_{1}}=
\sum_{k=0}^{2m-1}Tr_{R}\left({\ql^{2}\over {2r}}\Big({rk}- 2\ql\Big)(-1)^{
F_R}\right)\quad {\rm mod}\ 2mr$$

Now, you might think that We should demand
that each of these anomaly coefficients vanish ($mod\ 2mr$) separately. In
fact, this is too strong, and is not even true for many (2,2) models. Instead,
because of the Green-Schwarz mechanism, we can cancel the anomaly by
assigning a nontrivial transformation under $\gamma$ to the axion. But this
only can succeed in canceling all three anomalies provided the respective
anomaly coefficients obey the relations \DKthree
\eqn\anomrels{\left.
\matrix{
{\rm A}_1-{\rm A}_2&=0\cr
{\rm A}_3-24 {\rm A_1}&=0\cr
{\rm A}_3-24 {\rm A_1}&=0\cr}\right\} mod\ 2mr}
In particular, this means that the anomaly must vanish for any field
configuration satisfying $\Tr R^2=\Tr F_1^2+\Tr F_2^2$.

For most of the models we discussed in \S4, the anomaly in the ``quantum"
discrete R-symmetry does indeed cancel.  But, as alluded to earlier, some
models can be ruled out on this basis.

\newsec{$\bar Q^+$ Cohomology}

Finally, let us get down to constructing the spectrum of spacetime
fermions. We work on the cylinder, and as I said before, the states with
$\bar L_0^{(int)}=0$ can be represented as elements of the cohomology of one
of the right-moving supercharges
$\bar Q^\pm$, which we will take to be $\bar
Q^+$. The $\bar Q^+$ operator has three terms:
\eqn\barQplusdef{\eqalign{
\bar Q^+&=\bar  Q^+_R+\bar Q^+_L+ \bar Q^+_{spectator}\cr
\bar  Q^+_R&=\oint i\bar\psi_i\delb\phi_i\cr
\bar Q^+_L&=\oint \gamma^\alpha W_\alpha(\phi)+\lambda^a F_a(\phi)\cr
\bar Q^+_{spectator}&=\oint i\bar \psi_s \delb s+ \xi s\cr}}
The contribution to the supercharge from the spectator fields, $\bar
Q^+_{spectator}$, is decoupled from the rest of the $\bar Q^+$ operator, and
is purely quadratic. This means that it has trivial cohomology, and all the
physical states have representatives in which the $\Xi,S$ oscillators are in
their ground states. (This may, to the reader, seem like a fancy way of
noting that since $\Xi,S$ are massive, they cannot contribute to the states
of the IR conformal field theory, but it is nice to see the same result
emerging from this point of view as well.)

In any case, we need to concentrate on the cohomology of $\bar Q^+_R+\bar
Q^+_L$. Again, we note that $\bar Q^+_R$ is quadratic, and so has trivial
cohomology. However, it is coupled to $\bar Q^+_L$. Still, a standard
spectral sequence argument \rBott\ says that we can construct the
cohomology of
$\bar Q^+_R+\bar
Q^+_L$ starting with the cohomology of $\bar Q^+_R$ as a ``first
approximation". Actually applying the spectral sequence argument is a little
delicate here, because the complex is infinite-dimensional. However, the
result is basically correct \Us, and the cohomology of  $\bar Q^+$ is
\eqn\spectres{{\rm H}^*_{\bar Q^+}={\rm H}^*_{\bar Q^+_L}({\rm H}^*_{\bar
Q^+_R}\cap {\rm H}^*_{\bar Q^+_{spectator}})}

The cohomology of $\bar Q^+_R$ is trivial to calculate.
$$\eqalign{{\rm H}^*_{\bar
Q^+_R}=&\hbox{independent of}\ \psi_i,\bar\psi_i\ \hbox{oscillators and}\cr
&\hbox{(in untwisted sector) holomorphic in the zero mode of}\ \phi_i\cr}$$

We simply have to compute the cohomology of $\bar Q^+_L$ on this smaller
Hilbert space. This is made easier by the fact that, for these purposes, we
can use free-field OPEs -- the corrections from including the effect of the
superpotential interactions are $\bar Q^+$-trivial \WitMin. So, more or less,
we have reduced the problem to a free-field orbifold calculation, where the
boundary conditions on the fields are twisted by the action of $g^k$, where
$g$ is given by \gsoLGdef.

As usual in orbifold calculations, we expand the fields in oscillators
which, because the boundary conditions are twisted by $g^k$, are fractionally
moded. Also, the ground states of the twisted sectors carry fractional fermion
number, and hence fractional values of the charges
$\ql,\qr$. In our case \refs{\Us, \DK}\foot{Note that in Table 5, and
here, we have made a slight change in our notation for the $U(1)$ charges
of the fields from that of
\DK.},
\eqn\efraccharge{\eqalign{
\ql&=q_i\theta_{ik}+q_a\theta_{ak}+q_\alpha\theta_{\alpha k}\cr
\qr&=\bar q_i\theta_{ik}+\bar q_a\theta_{ak}+\bar q_\alpha\theta_{\alpha k}\cr
}}
where
\eqn\thetadef{\eqalign{
\theta_{ik}&={k (q_i-1)\over 2}+\left[{k (1-q_i)\over 2}\right] +{1\over 2}\cr
\theta_{ak}&={k (q_a-1)\over 2}+\left[{k (1-q_a)\over 2}\right] +{1\over 2}\cr
\theta_{\alpha k}&={k (q_\alpha-1)\over 2}+\left[{k (1-q_\alpha)\over 2}\right]
+{1\over 2}\cr }}
Similarly, one calculates the ground state energy  (eigenvalue of $L_0-c/24$)
of the $k^{th}$ twisted sector. The answer depends on whether $k$ is even,
which gets paired with the R-sector for the $(16-2r)$ free left-moving
fermions, or $k$ is odd, which gets paired with the NS-sector for
the $(16-2r)$ free left-moving
fermions
\eqn\Egs{\eqalign{
E_{NS}&=-{5\over 8}+\sum {\theta_{a,k}^2\over 2}+\sum
{\theta_{\alpha,k}^2\over 2}-\sum {\theta_{i,k}^2\over 2}\cr
E_{R}&=-{(r-3)\over 8}+\sum {\theta_{a,k}^2\over 2}+\sum
{\theta_{\alpha,k}^2\over 2}-\sum {\theta_{i,k}^2\over 2}\cr
}}

{}From here, constructing the spectrum is straightforward, but a little boring.
One simply goes through , sector by sector, and computes the cohomology of
$\bar Q^+_L$ on the twisted fock space. The spacetime quantum numbers of the
states that we find are correlated with the $\qr$ charge. For the massless
states,
$$\eqalign{
\qr&=-1/2\qquad\hbox{right-handed spacetime fermions ($\in$chiral
multiplet)}\cr
\qr&=+1/2\qquad\hbox{left-handed spacetime fermions ($\in$antichiral
multiplet)}\cr
\qr&=+3/2\qquad\hbox{right-handed gauginos ($\in$vector
multiplet)}\cr
\qr&=-3/2\qquad\hbox{left-handed spacetime fermions ($\in$vector
multiplet)}\cr
}$$

The left spectral flow maps us between adjoining twisted sectors. So, {\it
e.g.}~the left-handed gauginos are assembled, as in Table 1, from the ground
state of the $k=0$ sector, $\ket{k=0}$, (a spinor of $SO(16-2r)$), two states
from the $k=1$ sector:
$$\eqalign{
\lambda^I_{-1/2}\lambda^J_{-1/2}&\ket{k=1}\cr
\left[\sum q_i\phi^i_{-q_i/2}\bar\phi^i_{-1+q_i/2}-\sum
q_a\lambda^a_{q_a/2}\bar\lambda^a_{-1-q_a/2}-\sum
q_\alpha\gamma^\alpha_{q_\alpha/2}\bar\gamma^\alpha_{-1-q_\alpha/2}\right]
&\ket{k=1}\cr
}$$
and the ground state of the $k=2$ sector, $\ket{k=0}$, (again a spinor of
$SO(16-2r)$). Each of these is evidently annihilated by $\bar
Q^+_L=\oint(\gamma^\alpha W_\alpha(\phi) \lambda^a F_a(\phi) )$, and cannot
be written as $\bar Q^+_L$ of some other state.

 Similarly, we can go through
and construct the rest of the states in the spectrum of massless  fermions.
Some examples are worked out in detail in
\DK.

As an exercise, I recommend that the reader work out the $\bar Q^+$
cohomology for the model which, in the \cy\ phase, corresponds to a complete
intersection of two sextics in $W\BP^5_{1,1,2,2,3,3}$ with the left-moving
fermions coupling to a certain rank-4  vector bundle on it. The
$U(1)$ charges of the fields in the linear \sm\ (including the spectators) are
listed
in Table 6.

\hbox{\vtop{\hsize=2.25in
\def\tablerule{\omit&
\multispan{4}{\tabskip=0pt\hrulefill}&\cr}
\def\tablepad{\omit&
height3pt&&&&\cr}
$$\vbox{\offinterlineskip\tabskip=0pt\halign{
\hskip.5in
\strut$#$\quad&
\vrule#&\quad\hfil $#$ \hfil\quad &\vrule #&\quad
\hfil $#$ \quad&\vrule #\cr
&\omit&\hbox{Field}&\omit&Q&\omit\cr
\tablerule\tablepad
&& \Phi^{1,2} &&1&\cr
\tablepad\tablerule\tablepad
&& \Phi^{3,4} &&2&\cr
\tablepad\tablerule\tablepad
&& \Phi^{5,6} &&3&\cr
\tablepad\tablerule\tablepad
&&P&&-8&\cr
\tablepad\tablerule\tablepad
&&\Lambda^{1,2,3,4}&&1&\cr
\tablepad\tablerule\tablepad
&&\Lambda^{5}&&4&\cr
\tablepad\tablerule\tablepad
&&\Gamma^{1,2}&&-6&\cr
\tablepad\tablerule\tablepad
&&S&&-4&\cr
\tablepad\tablerule\tablepad
&&\Xi&&4&\cr
\tablepad\tablerule
\noalign{\bigskip}
\noalign{\narrower\noindent{\bf Table 6:}
$U(1)$ charges of the (bosonic and fermionic) chiral superfields in the
$W\BP^5_{1,1,2,2,3,3}$ models.}
 }}$$}
\vtop{\advance\hsize by -2.25in
Since $m=8$, there are 16 twisted sectors. You need not go through all of
them, as the states you find in sectors 9-15 are simply the CTP conjugates of
the states in sectors 1-7. You should find, for instance, that the
right-handed {\bf 16}s of $SO(10)$ arise in this model as states in the $k=0$
and $k=1$ sectors, corresponding respectively to the $8^s_{-1}$ and $8^v_{1}$
of
$SO(8)\times U(1)$. Explicitly, these states are
given by octic polynomials:
$$\eqalign{P_8(\phi^i_0)&\ket{k=0}\cr
\lambda^I_{-1/2}P_8(\phi^i_{-Q_i/16})&\ket{k=1}\cr}$$
For these to be in the $\bar Q^+_L$ cohomology, we must mod out the space of
octic polynomials by the ideal generated by the $F_a(\phi)$ and the
$W_\alpha(\phi)$. This yields 74 states in the $\bar Q^+_L$ cohomology.}}

\leftline{\bf Interactions}

We can also compute (unnormalized) Yukawa couplings of these fermions by
sandwiching an operator (the vertex operator
for a physical spacetime boson in a chiral multiplet), which commutes with
$\bar Q^+$, between two of the fermion states we have just constructed. The
results, at least in certain cases,
\refs{\DKfour,\TKawai} agree with those computed at large radius in the sigma
model, leading one to hope that there might actually be a theorem, analogous
to the one which holds in (2,2) theories \Exact, which states that these
couplings are independent of the \Ka\ moduli.

\newsec{Envoi}
Much, clearly, remains to be explored here. What are the analogues of
mirror symmetry? Can one compute the spacetime \Ka\ potential for the
fields? Is there indeed a theorem along the lines of \Exact\ for some of the
Yukawa couplings in these (0,2) theories? And what about the corrections to
those couplings which do get corrected?  Can one compute them in the linear
sigma model, along the lines of \MorrisonPlesser? Are there  exactly-soluble
conformal field theories (the analogues, perhaps, of the Kazama-Suzuki models
\KazamaSuzuki\ for (2,2) theories) which lie at special points in the moduli
spaces of (0,2) theories that we have been exploring? These, and many other
questions are waiting  to be answered.

\bigbreak\bigskip\bigskip\centerline{{\bf Acknowledgments}}\nobreak
\frenchspacing{
This work was supported by Robert A. Welch Foundation, NSF Grant PHY90-09850
and by an A.P. Sloan Foundation Fellowship. Much of the work described herein
was done jointly with S.~Kachru. I also benefitted from numerous discussions
with D. Morrison, R. Plesser, E. Silverstein and E. Witten.  }

\listrefs
\end